\documentclass[twocolumn]{aastex62}
\newcommand{\msun}{{\cal M_\odot}}
\newcommand{\boo}{{Boo~I}}
\graphicspath{{./}{figures/}}
\submitjournal{{AAS Journals}}

\shorttitle{Faint Stars in a Faint Galaxy: I.}
\shortauthors{Filion et al.}

\begin{document}

\title{Faint Stars in a Faint Galaxy:  I. Ultra Deep Photometry of the Bo{\"o}tes~I Ultra Faint Dwarf Galaxy}

\correspondingauthor{Carrie Filion}\email{cfilion@jhu.edu}
\author[0000-0001-5522-5029]{Carrie Filion}
\affil{{Department of Physics \& Astronomy, The} Johns Hopkins University, {Baltimore, MD 21218}}

\author{Vera Kozhurina-Platais}
\affiliation{Space Telescope Science Institute, {Baltimore, MD 21218}}

\author[0000-0001-9364-5577]{Roberto Avila}
\affiliation{Space Telescope Science Institute, {Baltimore, MD 21218}} 

\author{Imants Platais}
\affiliation{{Department of Physics \& Astronomy, The} Johns Hopkins University, {Baltimore, MD 21218}}

\author[0000-0002-4013-1799]{Rosemary F.G.~Wyse}
\affiliation{{Department of Physics \& Astronomy, The} Johns Hopkins University, {Baltimore, MD 21218}}



\begin{abstract}
  We present an analysis of new extremely deep images  of the resolved stellar population of the Bo{\"o}tes~I ultra faint dwarf spheroidal galaxy. These new data were taken with the Hubble Space Telescope, using the Advanced Camera for Surveys (Wide Field Camera) and Wide Field Camera 3 (UVIS), with filters F606W and F814W (essentially V and I),  as part of  a program to derive the  low-mass stellar initial mass function in this galaxy.  We {compare and contrast} two approaches to obtaining the stellar photometry, namely \textit{ePSF} and DAOPHOT. We identify likely members of Bo{\"o}tes~I {based on}  the location of each star on the color-magnitude diagram, obtained with  the DAOPHOT photometry from the ACS/WFC data. The probable members lie close to  stellar isochrones that were chosen to encompass the known metallicity distribution derived from spectroscopic data of brighter radial-velocity member stars and are consistent with the main-sequence turnoff.  The resulting luminosity function of the  Bo{\"o}tes~I galaxy has a 50\% completeness limit of 27.4 in F814W and  28.2 in F606W (Vega magnitude system), which corresponds to a limiting stellar mass of $\simlt 0.3 {\cal M}_\odot$.

\end{abstract}



\section{Introduction} \label{sec:intro}
The stellar initial mass function (IMF) describes the distribution of
stellar masses formed in any single star formation event {
  \citep{salpeter}.  Determinations} of the IMF for star clusters and
field stellar populations of the Milky Way {Galaxy}, over a range of ages and
metallicities, have found {the IMF to be} consistent with being
invariant, {and  well fit by a broken power law or a lognormal distribution (see, for example, the recent analysis of \citealt{IMF2019} and the reviews of \citealt{kroupa-review1}, \citealt{chabrier}, \citealt{bastian} and \citealt{kroupa-review2}).} Whether or not the IMF for
other galaxies is the same as that of the Milky Way has been the
subject of {much recent} debate {(cf. \citealt{bastian})}. Understanding {if, and how,} the IMF changes is crucial
{to many areas of astrophysics, from  stellar-mass estimates of high-redshift galaxies, to models of stellar feedback in simulations of galaxy formation.} 
The nearby low surface-brightness satellite galaxies of the Milky Way are obvious targets for study of the IMF, particularly since their low mean metallicity and high inferred dark-matter content provides an environment very different from the local disk. Galaxies whose stellar populations have a narrow range of ages offer the cleanest case studies. The luminosity function of the lower main sequence, below the old(est) turnoff, can be transformed into the IMF, modulo {corrections for} unresolved binary {systems, possible mass-dependent dynamical effects and observational} incompleteness. {The high-mass IMF in self-enriched systems can be constrained by the elemental abundances of the most metal-poor stars (cf.~\citealt{wg92,wyse98}).  

  \cite{grillmair}, in a pioneering study, obtained imaging data with the Hubble Space Telescope (HST) that reached 3 magnitudes below the old main-sequence turnoff in the Draco dwarf spheroidal (distance $\sim 80$~kpc). They determined the luminosity function of stars in Draco around the turn-off region, corresponding to a stellar mass range of $ 0.6 \simlt {\cal M/M_\odot} \simlt 0.9$, and found that it was very similar to that in the old, metal-poor {Milky Way} globular cluster M68. They also found that their data were consistent with  an underlying power-law  IMF, with slope similar to that of the IMF derived for stars in the solar neighborhood.  \citet{feltzing99} obtained deeper data with HST for the somewhat closer, but similarly old and metal-poor, Ursa Minor dSph (hereafter UMi, distance $\sim 66$~kpc). Their color-magnitude data extend more than 4 magnitudes below the main-sequence turnoff and, similarly as found for Draco, the luminosity function of the stars in UMi was found to be indistinguishable from that measured for an old, metal-poor {Milky Way} globular cluster (M92), in this case down to a magnitude corresponding to  mass $\sim 0.45 {\cal M_\odot}$. The subsequent analysis of the full deep HST data for UMi, in  \cite{wyse}, extended the luminosity function to reach a corresponding mass $\sim 0.3 {\cal M_\odot}$ and again found excellent agreement between  the luminosity functions of UMi and  old, metal-poor {Milky Way} globular clusters (M92 and M15). The age and (mean) metallicity of the stellar population in UMi are the same as those of M92 and M15, so that provided internal and external dynamical effects may be neglected for the globular cluster data (as argued by those authors), a comparison of the luminosity functions  is equivalent to a comparison of the IMFs. The inferred IMF for the Draco dSph and that for the UMi dSph are also in good agreement, over the limited mass range in common. 

The discovery of Ultra Faint Dwarf galaxies {(UFDs, galaxies with $L < 10^5 L_\odot$, see reviews of \citealt{2013review,simon-review})} allowed the exploration of the IMF in even lower luminosity, more dark-matter dominated systems.}   \cite{geha} {targeted two UFDs, Hercules and Leo IV, at distances of 135~pc and 156~kpc, respectively. The data allowed investigation of the IMF in each UFD over the mass range  $ 0.5 \simlt {\cal M/M_\odot} \simlt 0.8$. Comparison with the published low-mass IMF slopes for the three other galaxies with deep star-count data (the Milky Way, the Small Magellanic Cloud and UMi) led \cite{geha} to conclude that the low-mass end of the IMF varied systematically across the sample, in the sense that the galaxies with lower metallicity, or with lower stellar velocity dispersion, had a more bottom-light IMF. \cite{gennaro} reanalyzed the data for the two UFDs and expanded the sample  with the addition of  4 more (Coma Berenices, Ursa Major~I, Canes Venatici~II, and Bo{\"o}tes~I - hereafter \boo). Their results provided more evidence for a bottom-light low-mass IMF in UFDs compared to the Milky Way, albeit that the quoted  significance of the variation depended on the adopted form of the IMF  (typically $\sim 95$\% for a power-law, $\sim 68$\% for a log-normal mass function).

{Indeed, the robustness  of any conclusions as to the variability of the IMF with these data, which cover only the mass range $0.5 \simlt {\cal M}/\msun \simlt 0.8 $ (the upper limit being set by the main-sequence turn-off), was investigated by \citet{elbadry}. Those authors concluded that it would be necessary to obtain data that reached down to masses below $\sim 0.3 {\msun}$, a limit met by only the UMi dataset of \cite{wyse}, for the extragalactic systems studied thus far and discussed above.}

{This paper is the first in a series describing the results of our  program to determine the low-mass IMF down to $\simlt 0.3 \msun$ in  the UFD \boo\ and revisit the issue of  possible variability of the IMF. This present paper focuses on the  more technical aspects of the data analysis, as we aim to push the  luminosity function as faint as possible.  We begin, in Section~\ref{sec:boo},  with a brief motivation for the choice of \boo. We  then (Section~\ref{sec:observations})  describe the new observational data, taken with the Hubble Space Telescope (HST) under program GO-15317 (PI I.~Platais), together  with archival data that we use as an off-field. We adopted two approaches to the photometric reductions (\textit{ePSF} and DAOPHOT) {and illustrate the relative strengths of each}, as described in Section~\ref{sec:photandastACS}. {Our science goal requires pushing the data to their faintest limits, which in turn requires a thorough understanding of possible systematics. We therefore document in some depth the detailed procedures we used.} Our approach to the determination of the likelihood of membership of \boo\ for the stars in our sample is presented in Section~\ref{sec:CMD}, together with the resulting cleaned color-magnitude diagram of likely members.  We close, in  Section~\ref{sec:discussion}, with a brief discussion of the achieved lower-mass limit and  {the next steps in the later papers in this series.}}

\section{{Bo{\"o}tes~I} Ultra Faint Dwarf Galaxy}\label{sec:boo}

{\boo\ was discovered by \citet{bootesdiscovery} in the Sloan Digital Sky Survey imaging data. This system is relatively luminous, with $L_V \sim 3 \times 10^4 L_\odot$ \citep{martin,okamoto} and is  at a distance of  $\simlt 65$~kpc, comparable to the distance of UMi, and thus is significantly closer than the other UFDs of comparable luminosity studied by \citet{geha} and \citet{gennaro}, which are located  beyond 100~kpc. Such a combination of brightness and closeness is beneficial for pushing the stellar luminosity function to the required faint limits. The previous HST observations of \boo\ (GO-12549, P.I. T. Brown), used in \cite{gennaro}, reached an approximate fifty percent completenes limit at magnitudes corresponding to $\sim 0.5\msun$ and our intention is to reach $\sim 0.3 \msun$.}

The stellar population of \boo\ is old and metal-poor and its color-magnitude diagram is very similar to that of the globular cluster M92 \citep{bootesdiscovery}.  \boo\ has been the target of several spectroscopic and imaging surveys and its basic properties are given in Table~\ref{tab:boo-param}.  The wide-area imaging surveys  found that \boo\ is somewhat} elongated, with an ellipticity {$\epsilon \sim 0.2 - 0.3$} (\citealt{bootesdiscovery}, \citealt{martin}, \citealt{okamoto}, \citealt{Roderick}, \citealt{munoz_boo}), and has  extended stellar substructure, most recently quantified by \cite{Roderick}. {The proper motion of \boo\ has been derived from \textit{Gaia} DR2 \citep{gaia}, with the most recent analysis \citep{simon} giving  $\mu_\alpha cos\delta = -0.472 \pm 0.046~\text{mas}\text{yr}^{-1}$, $\mu_\delta = -1.086 \pm 0.034 ~\text{mas}\text{yr}^{-1}$. The corresponding orbital motion has a pericenter of greater than 30~kpc (e.g.~$45\pm 5$~kpc; \citealp{simon}) which is large enough that Galactic tidal forces have probably not been important in sculpting the stellar content.

\begin{deluxetable}{cccccccc}
\tabletypesize{\footnotesize}
\tablecolumns{3}
\tablewidth{0pt}
\tablecaption{{Properties of \boo} \label{tab:boo-param}}
\tablehead{
\colhead{Parameter} & \colhead{Value} & \colhead{Reference}} 
\startdata
RA (J2000) & 14:00:06 & 1\\
Dec (J2000) & +14:30:00 & 1\\
$\ell$ & $358^\circ.1$ & 1\\
{\it b} & $69^\circ.6$ & 1\\
$A_V$ & 0.047 & 2 \\
$r_h$ (exponential) & $12\arcmin .8 \pm 0\arcmin .7$ & 3 \\
$\epsilon$ & 0.22 & 3\\
Position Angle & $14^\circ .2$ & 3 \\
$M_V$ & $ -5.92{\pm0.2}$ & 3\\
$(m-M)_0$ &  $18.96\pm0.12$ & 4\\
Distance & $62\pm4$ kpc & 4\\
$\langle \ {\rm [Fe/H]} \ \rangle$ & {$-2.55$} & 5, 6 \\
$\Delta{\rm [Fe/H]} $ & 1.70 & 5 \\
Age & $\sim 13$~Gyr & 1, 3, 7 \\ [1ex]
\enddata
\tablecomments{References: (1) \cite{bootesdiscovery}, (2) \cite{schlafly}, (3) \cite{okamoto}, (4) \cite{siegel}, (5) \cite{norris_2010b}, (6) \cite{gilmore},  (7) \cite{brown}}
\end{deluxetable}

Spectroscopic studies {have established that the iron-abundance distribution peaks at  ${\rm [Fe/H]} \sim -2.5$,} with a {metal-poor} tail extending to ${\rm [Fe/H]} \sim -3.5$ and a sharp decline {more metal-rich of the peak, to  ${\rm [Fe/H]} \sim -1.8$ (\citealt{norris_2010b}, see also  \citealt{norris_2008},  \citealt{lai}, \citealt{brown}).} Analyses {of samples with high-resolution spectroscopy, and thus}  abundances for individual elements, find a {range} in ${\rm  [\alpha/Fe]}$, from $\sim 0.1$ to $\sim 0.5$ \citep{feltzing} {consistent} with a modest trend of increasing ${\rm [\alpha/ Fe]}$ with decreasing [Fe/H] (\citealt{gilmore}, \citealt{ishigaki}, \citealt{frebel}). This trend suggests that the duration of star formation in \boo\ was long enough for gas to be enriched by Type~Ia supernovae, {as  discussed} in \cite{gilmore} and \cite{frebel}. The {delay-time distribution for Type~Ia supernovae is a topic of some debate and, in a further complication, sub-Chandrasekhar mass explosions have been implicated in the early iron enrichment of dSph \citep{subchandra}. The minimum time for the onset of the incorporation of iron from a double-degenerate  binary is expected to be $\simgt 10^8$~yr \citep[e.g.][]{tammy},  while the reported age distribution  of stars in \boo, estimated from the spread in color of the turn-off region, has 97\% with an age of 13.3~Gyr and 3\% with an age of 13.4~Gyr (on a scale with the age of M92 being 13.2Gyr; \citealp{brown}). We will revisit the possible age range of the stars in \boo\ in paper~III of this series. } 

The range of {chemical abundances apparent in the members of  \boo\ is expected to lead to a broadening, in color, of the main sequence, particularly} below  $\sim 0.5 \msun$, reflecting changes in {the dominant source of opacity} for these cool stars (as discussed in e.g. \citealt{allard} and \citealt{baraffe}). {This enhanced broadening for the lower main sequence is compounded, in the observational dataset presented here, by the larger photometric errors for fainter sources. Knowledge of the elemental abundance distributions from complementary data, such as for \boo, is an important element in the choice of target galaxy.}

%

\section{Observations}\label{sec:observations}
We obtained deep photometry of the \boo\  UFD, with the aim of pushing the faint magnitude limit below {that achieved in earlier datasets},  and thus extending the stellar mass regime available for study of the IMF. The new observations for this project were taken during twenty-six orbits of HST from May 30th to July 5th, 2019, utilizing the Advanced Camera for Surveys (ACS) as primary instrument and the Wide Field Camera 3 (WFC3) in parallel mode.  The ACS observations were taken with the Wide Field Camera (WFC), and the  WFC3 observations were taken with UVIS. For future reference, the field of view of ACS/WFC is $202 '' \times 202''$, with a resolution of $0.05 ~\text{arcsec} ~\text{pix}^{-1}$, while the field of view of WFC3/UVIS is $162~\arcsec \times 162\arcsec$, with  a resolution of $0.04~\text{arcsec}~\text{pix}^{-1}$. Twenty-four orbits were dedicated to on-target observations of \boo\ -- more than four times the exposure time of the previous deep observations of \boo\ --  with the remaining two orbits dedicated to an off-field. The data described here may be obtained from the MAST archive at \dataset[doi:10.17909/t9-jr7h-en65]{https://dx.doi.org/10.17909/t9-jr7h-en65}.

The on-field observations were taken in three pointings, the footprint of which is indicated in Figure~\ref{fig:footprint}. The ACS/WFC fields match three of those observed with ACS/WFC by  GO-12549; this allows for the possibility of a deep proper-motion study, the topic of a future paper. We also selected the same ACS/WFC filters as this earlier work, namely F606W and F814W, and the equivalent passbands for the parallel observations with WFC3/UVIS.  The ACS/WFC fields selected were pointings one, three, and five of GO-12549 and we retained their naming convention, so that pointing three encompasses the central regions of \boo\ (see Table~\ref{tab:obs} for the coordinates of the field centers). As detailed in Table~\ref{tab:obs}, each field of  \boo\ was imaged over four orbits in each filter, and two exposures were taken per orbit.  A four-point dither pattern was used to mitigate the effects of the under-sampled point spread function (PSF)  of the instruments on HST (see \citealp{lauer1,lauer2} for more discussion). The dithering strategy also optimized the rejection of detector artifacts and cosmic rays. {One of the two guide stars used to obtain fine-lock guiding of the telescope could not be acquired for orbit~10, during observations of pointing one.   The availability of only one guide star  resulted in some loss of astrometric accuracy during this orbit, but happily there was} no degradation in image quality. 

\begin{figure}[ht]
\begin{center}

\includegraphics[width=0.5\textwidth]{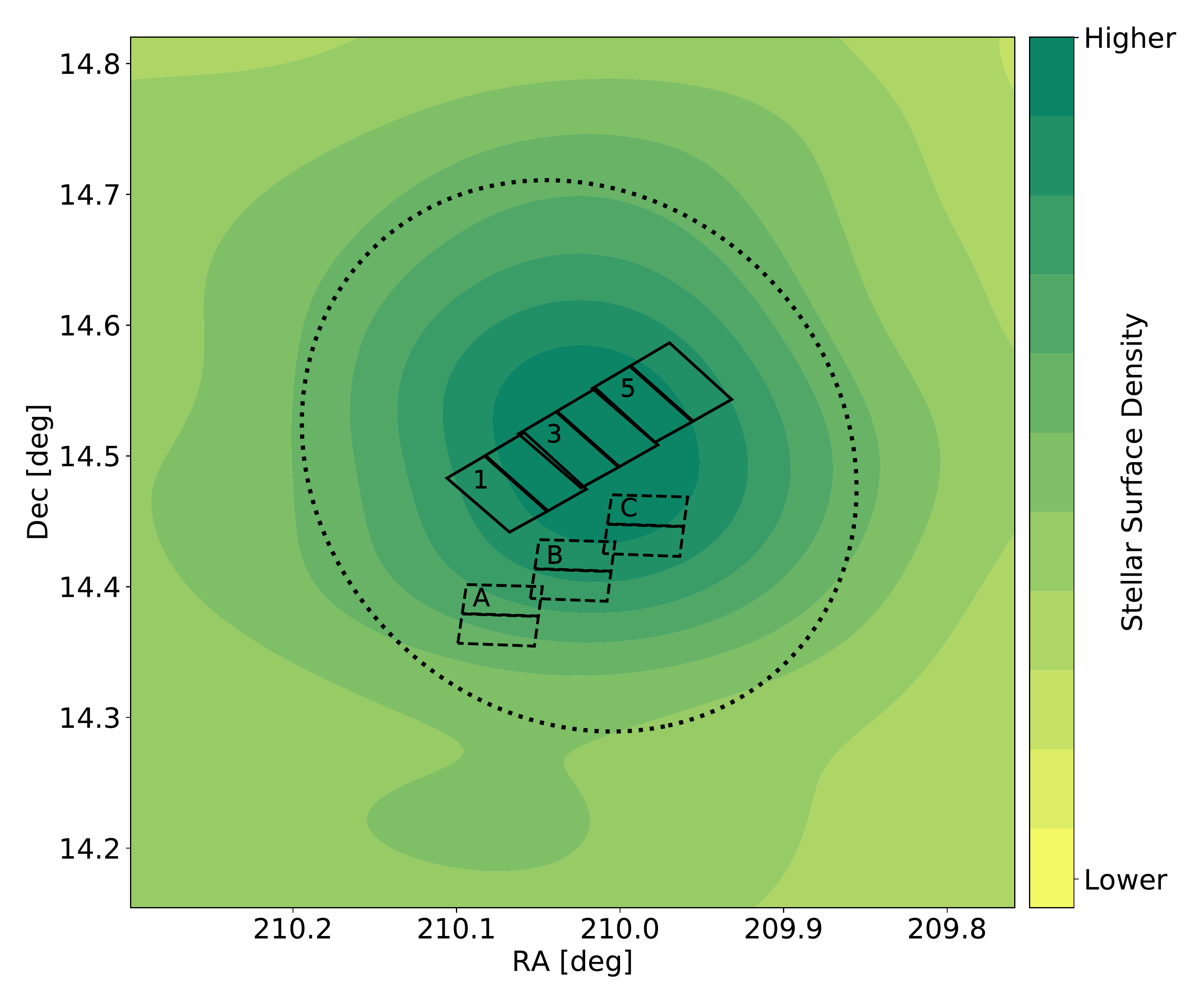}

\caption{The footprint of the observed fields of \boo.   The solid rectangles closest to the center represent the ACS/WFC fields, and the dotted rectangles represent the parallel WFC3/UVIS fields. The naming schema of our pointings is East to West, i.e. pointing~1 of the ACS/WFC observations and pointing~A of the WFC3/UVIS observations are Eastmost. The green contours indicate the {stellar surface density}, which we created {from the photometry published in \cite{munoz_description}. The  dashed oval represents the half-light (elliptical) radius from  \citet{okamoto}, corresponding to $\sim 230$~pc at an assumed distance of 62~kpc.}}

\label{fig:footprint}

\end{center}
\end{figure}

The off-field was imaged {for one orbit in} each filter, with only one exposure. {The off-field data are intended  to aid in the quantification of the contribution of background unresolved galaxies and}  non-member stars {to the \boo\ fields. As discussed in Section~\ref{sec:offfield_phot}, {this one long exposure proved to be difficult to clean. We therefore searched the  MAST archive and identified a dataset to use as a} second off-field, with Galactic coordinates similar to those of \boo, and with {multiple-exposure} ACS/WFC imaging in the same filters (designated `Off Field 2' in Table~\ref{tab:obs})}.

{As noted in Section~\ref{sec:intro}, our aim of determining the low-mass stellar IMF places stringent requirements on the data analysis, and the following two sections  are, by necessity,  technical. Readers who  are less interested, or who are very familiar with the photometric reduction techniques, may prefer to turn to Section~\ref{sec:CMD} where the resultant deep color-magnitude diagrams are presented.}

\begin{table*}
\caption{{Details of the observations}}
\label{tab:obs}
\begin{center} 
\begin{tabular}{l c c c c c c c r}
  \hline
  \hline 
Pointing & Instrument & Filter & $\alpha_{2000}$  & $\delta_{2000}$  & Start UT Date & $N_{exp}$  & Total Exp. Time \\  
&  &  & [hh:mm:ss] & [$^{\circ}$:\arcmin:\arcsec] & [yymmdd] &  & [seconds]  & \\ [2ex]
\hline 

Pointing 1 & ACS/WFC & F606W & 14:00:15.510 & +14:28:42.06 &  190530 & 8 & 9792 \\	
Pointing 1 & ACS/WFC & F814W & 14:00:15.510 & +14:28:42.06 &  190530 & 8 & 10064 \\ 
Pointing 3 & ACS/WFC & F606W & 14:00:04.680 & +14:30:47.17 &  190630 & 8 & 9792 \\ 
Pointing 3 & ACS/WFC & F814W & 14:00:04.680 & +14:30:47.17 &  190630 & 8 & 10064 \\
Pointing 5 & ACS/WFC & F606W & 13:59:53.847 & +14:32:52.28 &  190610 & 8 & 9792 \\
Pointing 5 & ACS/WFC & F814W & 13:59:53.847 & +14:32:52.28 &  190610 & 8 & 10064 \\
Off Field & ACS/WFC & F606W & 13:59:54.030 & +15:57:45.65 &  190619 & 1 & 2603 \\
Off Field & ACS/WFC & F814W & 13:59:54.030 & +15:57:45.65 &  190619 & 1 & 2660 \\
Pointing A & WFC3/UVIS & F606W & 14:00:17.882 & +14:22:49.28 & 190530 & 8 & 10112 \\
Pointing A & WFC3/UVIS & F814W & 14:00:17.882 & +14:22:49.28 & 190530 & 8 & 10312 \\
Pointing B & WFC3/UVIS & F606W & 14:00:07.053 & +14:24:54.39 & 190630 & 8 & 10112 \\
Pointing B & WFC3/UVIS & F814W & 14:00:07.053 & +14:24:54.39 & 190630 & 8 & 10312 \\
Pointing C & WFC3/UVIS & F606W & 13:59:56.221 & +14:26:59.50 & 190610 & 8 & 10112 \\
Pointing C & WFC3/UVIS & F814W & 13:59:56.221 & +14:26:59.50 & 190610 & 8 & 10312 \\
Off Field 2 & ACS/WFC & F606W & 14:10:38.005 & -11:43:58.62 &  140713 & 4 & 2080 \\
Off Field 2 & ACS/WFC & F814W & 14:10:38.005 & -11:43:58.62 & 140713 & 4 & 2110 \\ [2ex]
\hline
\end{tabular}
\\
\end{center}
\tablecomments{The naming convention of the ACS/WFC pointings follows that of GO-12549, so that pointing number increases with declination (i.e. pointing one has the lowest declination and pointing five has the highest). The WFC3/UVIS observations were taken in parallel mode, and the pointing name follows alphabetical order increasing with declination. The data for  Off Field 2 are archival and were obtained under program GO-13393.}
\end{table*}

\section{Photometry: ACS/WFC Data}\label{sec:photandastACS}

Photometry for the ACS/WFC on-field exposures was produced using two different techniques and software packages. The first, \textit{ePSF}, developed by \cite{anderson}, is an empirical approach that is able to map, with high fidelity, the changes in the true PSF across the CCD chip. \textit{ePSF} photometry can be performed only on individual exposures,\footnote{We use {\tt $\ast$\_flc.fits} files as the input into the photometry pipelines. These files are produced through the HST software CALACS, which subtracts the bias, flat-fields the image and corrects for charge transfer efficiency (CTE). Any time single exposures or single images are mentioned in this analysis, this refers to the {\tt $\ast$\_flc.fits} files.} which results in a brighter detection limit {than with co-added exposures}. The second photometry software {package that we employed} is {the stand-alone DAOPHOT II \citep{stetson} provided through Starlink \footnote{\url{http://www.star.bris.ac.uk/~mbt/daophot/}} \citep{starlink}}. DAOPHOT can be run on drizzle-combined images from multiple dithered exposures, which allows for a deeper detection limit. The single images for each pointing and filter were combined through \textsc{DrizzlePac}, {as described in detail in Section~\ref{sec:drizzle}.}  In contrast to the empirical \textit{ePSF}, the PSF fitting routine within DAOPHOT creates analytical models, which may not be flexible enough to capture the variation of the under-sampled PSF across the detector chip. We {performed} photometry using both \textit{ePSF} and DAOPHOT to {quantify how} well their results compare, and to understand what impact the reduced flexibility of DAOPHOT had on the final photometry. The resulting photometric catalogs are compared in Section \ref{sec:ePSFcomp}.

\subsection{\textit{ePSF} Photometry of ACS/WFC}\label{sec:ePSF}
\cite{anderson00} developed the effective PSF ({\it ePSF\/}) techniques to measure accurate stellar positions and magnitudes for HST WFPC2 images, where the PSF was severely under-sampled and varied across the CCD chips. This approach {was}  extended to ACS/WFC observations {by  \citet{anderson}.}  Further improvements include corrections for the effective area of each pixel (i.e. the pixel area map) that preserve the photometric accuracy \citep{acshandbook}. {A detailed description of the \textit{ePSF} technique may be found in \citet{anderson08a,anderson08b} and a short overview follows.}

{In the {\it ePSF\/} approach, the PSF is derived empirically by evaluation and scaling of the observed pixel intensity values.  No analytical function is fitted, in contrast to the PSF fitting methods used within DAOPHOT \citep{stetson}, DoPHOT \citep{dophot}, or DOLPHOT \citep{dolphot}. The} variations of the PSF across each detector chip is represented by an array of fiducial PSFs, between which each star's PSF is interpolated \citep{anderson}. The variation of the PSF with time, due {to, for example,} telescope breathing or small changes in focus, is presented within \textit{ePSF} as a spatially constant perturbation to the PSF for each image. A $5 \times 5$ pixel inner box is used to construct the \textit{ePSF} for the brightest pixels of each {detected point source, an area that} contain most of the flux. {The code returns, for each  individual exposure},  a list of precise measurements of the stellar {centroids, given as the X and Y positions in the detector coordinate system.  These are} then corrected for geometric {distortions.  The} flux in counts, ${\rm flux_{DN}}$, is converted into instrumental magnitudes by applying the standard expression: $2.5\log{(\rm flux_{DN})}$. {The code also returns a quality-of-fit parameter, \textit{qfit}, which} is used to separate stars with well-measured empirical PSFs from spurious detections, such as cosmic rays or hot pixels.

 \textit{ePSF} photometry  was {carried out} on each single exposure of the ACS/WFC fields.  Only measurements with $0<\textit{qfit}<0.5$ were selected for further examination. The {distortion-corrected}  X and Y positions from each exposure {in the F814W and F606W filters} were then matched to the positions from the central exposure {with respect to which the other exposures were dithered.} A matching tolerance {of} $<0.1$ pixels in {each of the X and Y coordinate} allowed {for the rejection of}  any remaining spurious detection in  the individual exposures. 

Finally, {three corrections} were applied to convert the photometric measurements from the individual images to the Vega magnitude system: \\
\begin{enumerate}
    \item an empirical aperture correction from the \textit{ePSF} radius {out}  to a radius of 10 pixels ($0.5\arcsec$ on the individual exposures). {The value of this correction was} found by calculating the difference between the \textit{ePSF} photometry and aperture photometry done using a {10-pixel radius, for isolated, well-exposed} stars on drizzle-combined images. This correction is small, ranging from $\sim 0.02$ to $\sim 0.03$ magnitudes in the different filters; 
    \item an aperture correction from 10 pixels to infinity, using the encircled energy estimate provided in \cite{bohlin};
    \item the filter-dependent Vega magnitude zero point for {the observation date of each pointing}\footnote{\url{http://www.stsci.edu/hst/instrumentation/acs/data-analysis/zeropoints}}
\end{enumerate} 

The final photometry for each star {was} calculated as a weighted average of the \textit{ePSF} measurements from each individual exposure. The photometric error in each filter {was estimated by} the root-mean-square deviation of the independent measurements for each star. The final photometric {catalog} for each pointing provides X and Y positions, based in the detector reference frame, and weighted photometry in the F606W and F814W filters. {We restricted the catalog to only sources that were present in at least three}  exposures per filter, which further {ensured high-quality} photometry. The resulting {color-magnitude} diagram from the \textit{ePSF} photometry for each individual pointing is presented in Figure~\ref{fig:epsf}. 

\begin{figure*}[ht]

\begin{center}

\includegraphics[width=\textwidth]{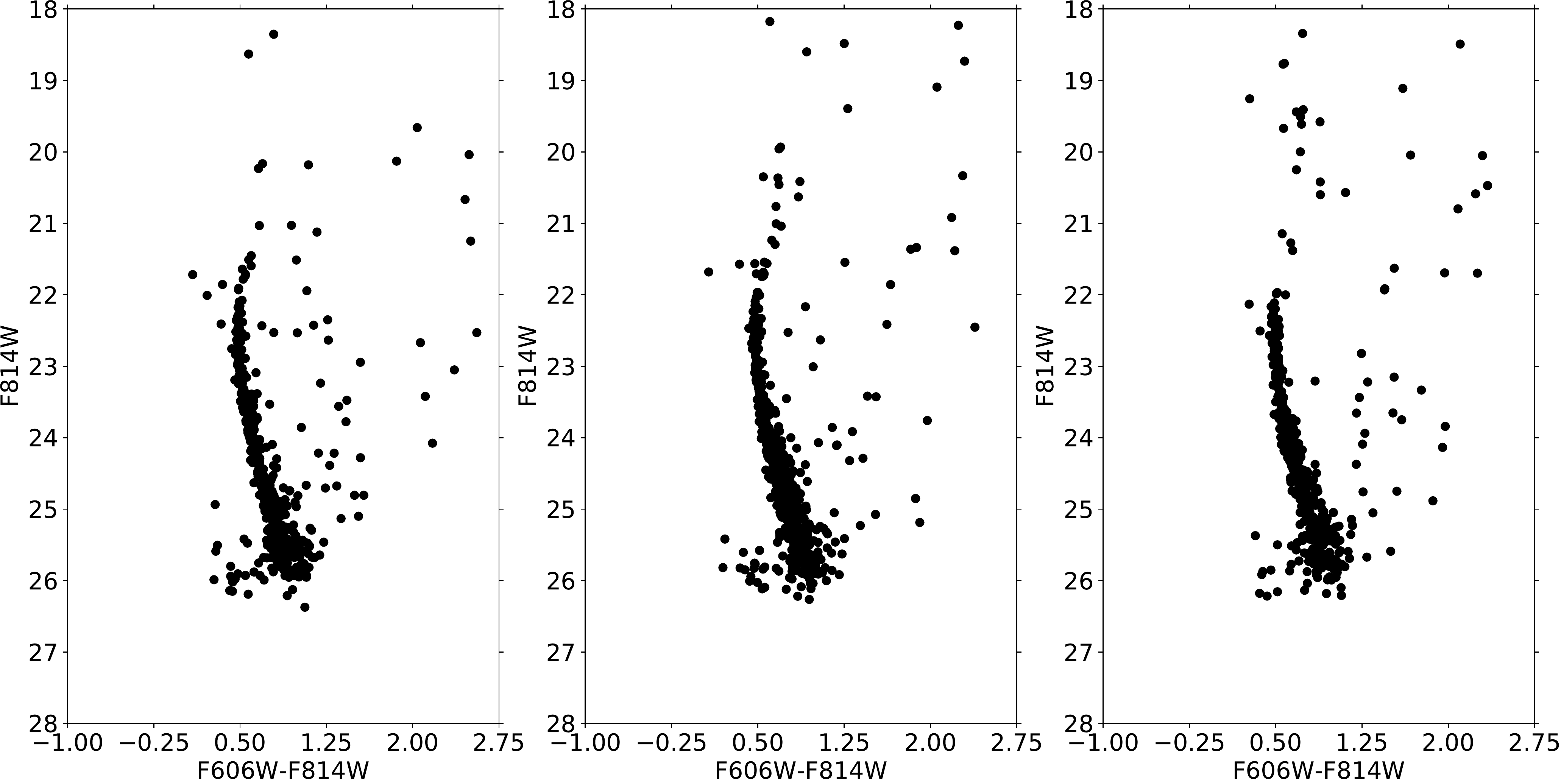}

\caption{Color-magnitude diagram {obtained from \textit{ePSF} for all three  on-field ACS/WFC pointings of our program. Only the  photometry for sources present in at least three}  exposures per filter is presented. The relative number of detections in each pointing are broadly consistent with what one would anticipate from the surface density contours seen in Figure \ref{fig:footprint}. Pointing one contains 484 sources and is leftmost, pointing three contains 588 sources and is central, and pointing five contains 501 sources and is rightmost. The main-sequence of \boo\  is {well delineated} down to a F814W  magnitude of $\sim 25$.}

\label{fig:epsf}

\end{center}

\end{figure*}

\subsection{DrizzlePac Aligning and Combining}\label{sec:drizzle}

{As noted above}, the PSF of HST instruments is under-sampled and varies spatially. The HST software \textsc{DrizzlePac} \citep{drizzle} was designed to rectify and combine dithered HST images into a common frame and, consequentially, enhance the spatial resolution and deepen the detection limit. {The} Mikulski Archive for Space Telescopes (MAST) provides users with { images for each field of a given program that have already been combined using \textsc{DrizzlePac}, adopting default values for the parameters of this package.} However, employing custom parameters (as described below) is recommended to exploit the full capabilities available through the drizzling algorithm, and improve accuracy of photometry. \textsc{DrizzlePac} also produces bad-pixel maps and cosmic-ray {masks, allowing for the identification of  problematic pixels that should be down-weighted during} the combining process. The final, combined images are cosmic-ray cleaned, {sky-subtracted}, undistorted, free from detector artifacts, and, depending on dithering pattern, free from chip gaps, as is the case {for the images analyzed in this paper.} These final drizzle-combined images also have improved resolution and depth compared to the input individual images. 

We {used \textsc{DrizzlePac} version 2.2.6}  {to distortion-correct}, align, and combine {the individual exposures.} The \textsc{DrizzlePac} task \textit{TweakReg} was used to {align the images and update} the WCS information in the headers of the input images.  \textit{TweakReg} finds sources in each image, corrects for distortion \citep{vera},
calculates shift, rotation and scale, and aligns to a specified reference image. All images {taken in} the same filter {at}  each pointing were aligned to one another utilizing the `general' (i.e.~six linear parameter) transformation. The F606W and F814W exposures were then aligned to each other, allowing for easier {subsequent} matching of detections between images.  The aligned {ACS/WFC} images  were then combined with {the \textsc{DrizzlePac} task}  \textit{AstroDrizzle}, adopting the parameter values \textit{final\textunderscore pixfrac} = 0.8 and  \textit{final\textunderscore scale} = $0.04\arcsec/$pixel. These parameters and their definitions are discussed in depth in the \textsc{DrizzlePac} resources provided by STScI\footnote{\url{http://www.stsci.edu/scientific-community/software/DrizzlePac.html}} and in \citet{gonzaga} and \citet{roberto}. {Briefly, the value of} \textit{final\textunderscore pixfrac} corresponds to the amount that each input pixel is shrunk before being added to the final pixel plane, where $ \textit{final\textunderscore pixfrac} = 0$ is equivalent to interlacing of pixels, and $\textit{final\textunderscore pixfrac}=1$ is equivalent {to shifting and adding} the pixels. The \textit{final\textunderscore scale} {parameter represents}  the scale of the output pixel, where the native ACS/WFC pixel scale is $\sim 0.05\arcsec/$pixel. {A side-by-side} comparison of a portion of a {distortion-corrected} individual ACS/WFC exposure (i.e.~the \lq\lq single drizzled'' image) from one pointing {with} the final, aligned and drizzle-combined image for the same pointing  {shown} in Figure~\ref{fig:exposureimage}. The {resolution has been noticeably improved, faint background galaxies are more distinguishable, and cosmic rays, hot pixels, and the chip gap were all} removed.  {The final step was to mask large, over-exposed stars and galaxies, achieved using \textit{imedit} within} IRAF \citep{iraf1,iraf2}.

\begin{figure*}[ht]

\begin{center}

\includegraphics[width=\textwidth]{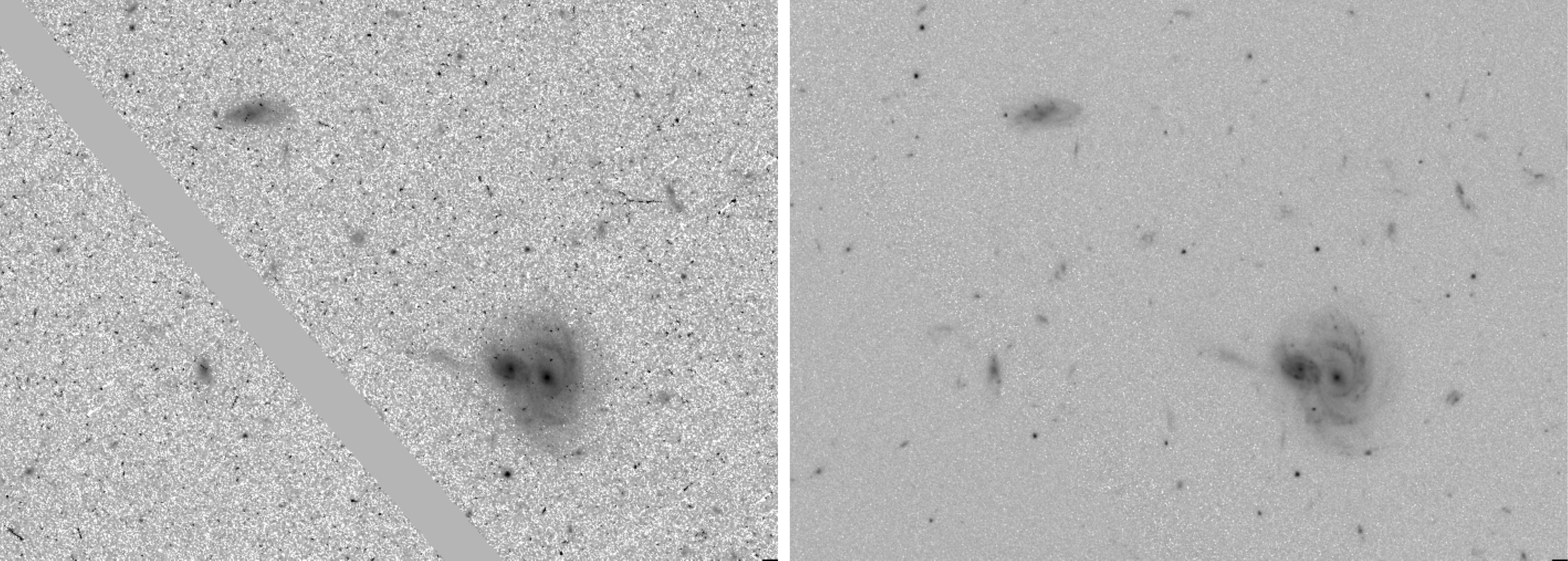}

\caption{A side-by-side comparison of a single drizzled image, left, with a drizzle-combined image, right, for pointing 5 (the westmost) of the ACS/WFC observations. These exposures are in filter F606W, with an individual exposure time of 1,224 seconds and a combined total exposure time of 9,792 seconds. Note the increased resolution in the drizzled product, as well as the uniform sky and clean removal of cosmic rays. The {region shown is} centered on $13:59:55.2, +14:33:27.0$ and is $\sim 45\arcsec \times 32 \arcsec$.}

\label{fig:exposureimage}

\end{center}

\end{figure*}
\subsubsection{Spatially Resolved Background Galaxies}

The data acquired in this program may be of interest for investigations of the stellar content of more distant galaxies than \boo.  The depth reached is intermediate  between that of the COSMOS HDF fields and that of the Ultra Deep Field (HUDF), albeit in only two filters. The morphology of intermediate-redshift, spatially resolved galaxies may be studied in detail.  For example, the left panel of Figure~\ref{fig:SDSS_comparison} shows a zoom-in on the prominent pair of spiral galaxies seen in Figure~\ref{fig:exposureimage}.  There is a wealth of internal structure in each galaxy, and possible tidal features.   The right panel shows a cutout of the same area of sky from the SDSS imaging data; the two galaxies are blended and only one  source is identified,   SDSS J135954.62+143321.8, with a reported  photometric redshift of $0.42$.  

\begin{figure*}[ht]

\begin{center}

\includegraphics[width=\textwidth]{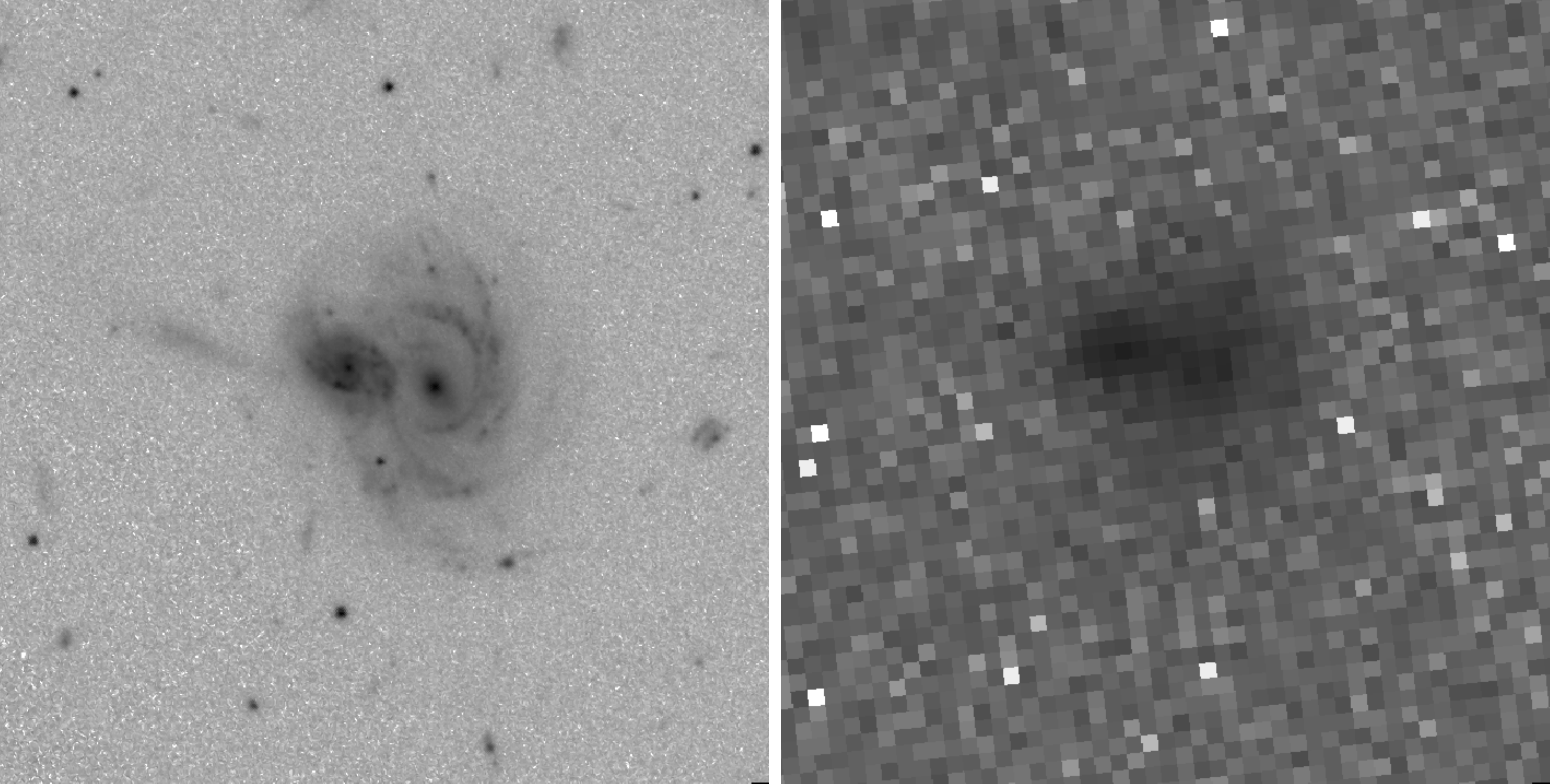}

\caption{{\it Left Panel:\/} A zoom-in on the spiral galaxies seen in the drizzle-combined image from Figure~\ref{fig:spatial_var} {\it Right Panel:\/}  The same galaxies in the SDSS imaging data.  The {internal} structure of both galaxies is resolved in the drizzle-combined image, with {spiral arms, central concentrations and possible tidal features all} evident.  {The SDSS image quality is insufficient to decompose the light into two galaxies and the source is identified as galaxy J135954.62+143321.8, at a photometric redshift of 0.42. The region shown is  centered on 13:59:54.69, +14:33:21.4, and is} $\sim 19\arcsec \times 15 \arcsec$.}

\label{fig:SDSS_comparison}

\end{center}

\end{figure*}

\subsection{DAOPHOT Photometry} \label{sec:DAOphot}
Aperture and PSF fitting photometry were performed with the DAOPHOT II package \citep{stetson}. The rationale for choosing DAOPHOT II {was} twofold, owing to {both} its strong performance {at detecting faint stars in  low signal-to-noise data and its}  ability to be run on drizzle-combined images. The spatial variation of the PSF can be modeled {within the DAOPHOT II package} up to a quadratic function. {The implementation of} DAOPHOT  on drizzled images {requires that the data} be in units of electrons {and that} the sky {background  be} added back in. {After such preparation of the ACS/WFC images for this project,} sources for photometry were {identified at a}   brightness threshold of four standard deviations above the sky background. These sources were then passed on to aperture photometry, and then to PSF fitting.

{The PSF in each pointing was modeled within DAOPHOT by fitting}  a Penny function  to well-exposed, isolated, stars, {adopting  a value of 10 pixels} (corresponding to $0.4\arcsec$ after drizzling) {for the `PSF radius' and 3 pixels for the  `fitting radius' (also referred to as the `inner radius').}  The Penny function \citep{penny} was chosen for its flexibility and success in fitting the PSF of our drizzle-combined images. {We tested all PSF functional forms available in DAOPHOT, and found that the Penny function produced the lowest $\chi$-squared goodness of fit value in the PSF fitting routine}. The PSF photometry was obtained with {the \textit{AllStar} routine within DAOPHOT. This routine first} fits the stars in the source list with a PSF, {then} subtracts those stars from the frame. The star-subtracted image was then run through the source finding, aperture photometry, and \textit{AllStar} routines again, to detect and PSF-fit fainter stars. {This iteration retrieved only   $\sim 12\%$ as many stars as were} retrieved in the first run {and this low incremental yield led us to truncate the process after two runs.  This decision was} supported by the results of the artificial star tests, as discussed in Section \ref{sec:artstar} {below}, which found a maximum of one additional (artificial) star in the second round of \textit{AllStar}.

The resulting {PSF-fitted} photometry was then converted to the Vega magnitude system, following a procedure similar to that used in the \textit{ePSF} {photometry, namely:}
\begin{enumerate}
    \item apply an aperture correction {out} to 10~pixels ($a_{3\rightarrow10}$), again using the difference between the aperture and PSF instrumental photometry for well-exposed stars. {The  DAOPHOT routine \textit{Aperture} was used} to determine the magnitude of each {given  star, within a 10-pixel aperture, in the} drizzle-combined images
    \item apply an aperture correction from ten pixels to infinity ($a_{10 \rightarrow \infty}$), from \cite{bohlin} 
    \item add the Vega magnitude zero point ($Z_p$) for the observation date and filter\footnote{\url{http://www.stsci.edu/hst/instrumentation/acs/data-analysis/zeropoints}}
    \item subtract the DAOPHOT zero point ($C$), which is equal to 25 \citep{stetson}
    \item {account for the exposure time ($t_e$) by adding $2.5\log(t_e)$, since} the drizzle-combined images are in units of electrons, not electrons per second 
\end{enumerate}
The final expression for magnitude conversion is {thus}: $m_{DAOPHOT} = m_{psf} - a_{3\rightarrow 10} - a_{10\rightarrow \inf} + Z_p - C + 2.5 \log({t_e})$, where $m_{psf}$ is the DAOPHOT instrumental PSF magnitude. {The error budget for each star is dominated by  the standard error provided by DAOPHOT and  the error bars shown in the color-magnitude plots below are calculated as the median of each of the standard error in color and magnitude,} in  magnitude bins of width $0.2$~mag. 

\subsection{Extinction Correction}\label{sec:dust}

We estimated the Galactic interstellar extinction using the dust maps from \cite{schlafly} and the ratio between extinction at the effective wavelength of our passbands and extinction in the Johnson V band \footnote{provided by \url{http://svo2.cab.inta-csic.es/theory/fps/index.php?mode=browse&gname=HST}}. {The reddening and extinction  varies less than 0.005~magnitudes across the full areal coverage  of the data and we adopted a uniform value of ${\rm A_V}= 0.047$,  equal to that for the nominal center\footnote{provided by \url{https://irsa.ipac.caltech.edu/applications/DUST/}} of \boo} and assumed that the reddening and extinction does not vary across our field of view. We did not correct for any possible extinction due to dust within \boo, {as significant amounts of dust are  unlikely to be present, given the low value of the upper limit on HI gas content of $< 1.8 \times 10^3 \msun$ \citep{putman}. The resultant} extinction correction is quite small, 0.04 magnitudes in F606W and 0.03 magnitudes in F814W. {Note that in all of the comparisons below between the photometric data and theoretical isochrones, it is the isochrones that have been shifted to account for the dust reddening and extinction.}

\subsection{Artificial Star Tests: Completeness and Cleaning}\label{sec:artstar}
Artificial star tests were {carried out using}  the DAOPHOT \textit{AddStar} routine. Three hundred {artificial stars, at brightnesses that randomly sampled the entire magnitude range of the data, were placed at random locations within}  each frame, and the resulting file was run through the  photometry pipeline. We did this thirty times per exposure per filter, {giving a total of} $54,000$ artificial stars added over 180 images. We chose to add {this number of stars (300) in order}  to avoid noticeably altering the {level of crowding, or overall surface density of sources, in the original exposures}, while keeping the number of new frames produced to a manageable size. The total number of artificial stars added per filter is $\sim 10 \times$ the number of stars in {the final, cleaned photometric catalog, above the 50\% completeness limit derived below (a total of 2,654 stars)}.

{The approach we used to clean  the data} of spurious, non-stellar detections that had made it through the photometry pipeline {is similar to that described by \cite{okamoto}}. The stars retrieved from the artificial star tests were used to establish the statistical properties that a perfect stellar PSF would have, after being retrieved through the pipeline. {Detected sources with  properties that fell too far from those of this ideal point source were then discarded. The properties that formed the basis for this  discrimination} between stars and artefacts were both the value of the DAOPHOT \textit{sharp} parameter, which describes the difference between the width of the object and the width of the PSF, {and the fractional error in the magnitude.}

{A true stellar source should have a \textit{sharp} parameter with value close to zero. The solid (red) curves in the lower panel of Figure~\ref{fig:cleaning} indicate the $\pm 3 \sigma$ bounds from the artificial star tests in the F606W images, derived from the mean and standard deviation computed in bins of 0.05~magnitudes. The fractional error on magnitude was obtained as}  the ratio of the standard error of the magnitude quoted by DAOPHOT and the instrumental, DAOPHOT PSF magnitude. {Again, the mean and standard deviation}  were computed within bins of 0.05 magnitude for the {artificial stars retrieved from the tests, and the solid (red) curve in the upper panel of Figure~\ref{fig:cleaning} represent the $+3\sigma$ bound. The real sources detected by DAOPHOT are represented by the points in both panels of Figure~\ref{fig:cleaning}.  Those sources that lie within $3\sigma$ of the mean defined by the artificial stars are denoted by the filled  circles and retained in the sample as likely stars, while sources that are rejected by one or other of these criteria (in each filter) are denoted by plus signs.  The sources that pass this cleaning process} have a mean and median goodness-of-fit parameter from DAOPHOT ($\chi$, related to the noise statistics) of approximately unity, as true stellar objects should.  A total of 3,035 stars pass through this step, out of an original 8,383 detections {and  2,654 stars have apparent magnitudes above the fifty percent completeness limit (derived below) in both filters. We present both of the resultant cleaned color-magnitude diagrams in Figure~\ref{fig:allcmd}.}

\begin{figure}[t]

\begin{center}

\includegraphics[width=.45\textwidth]{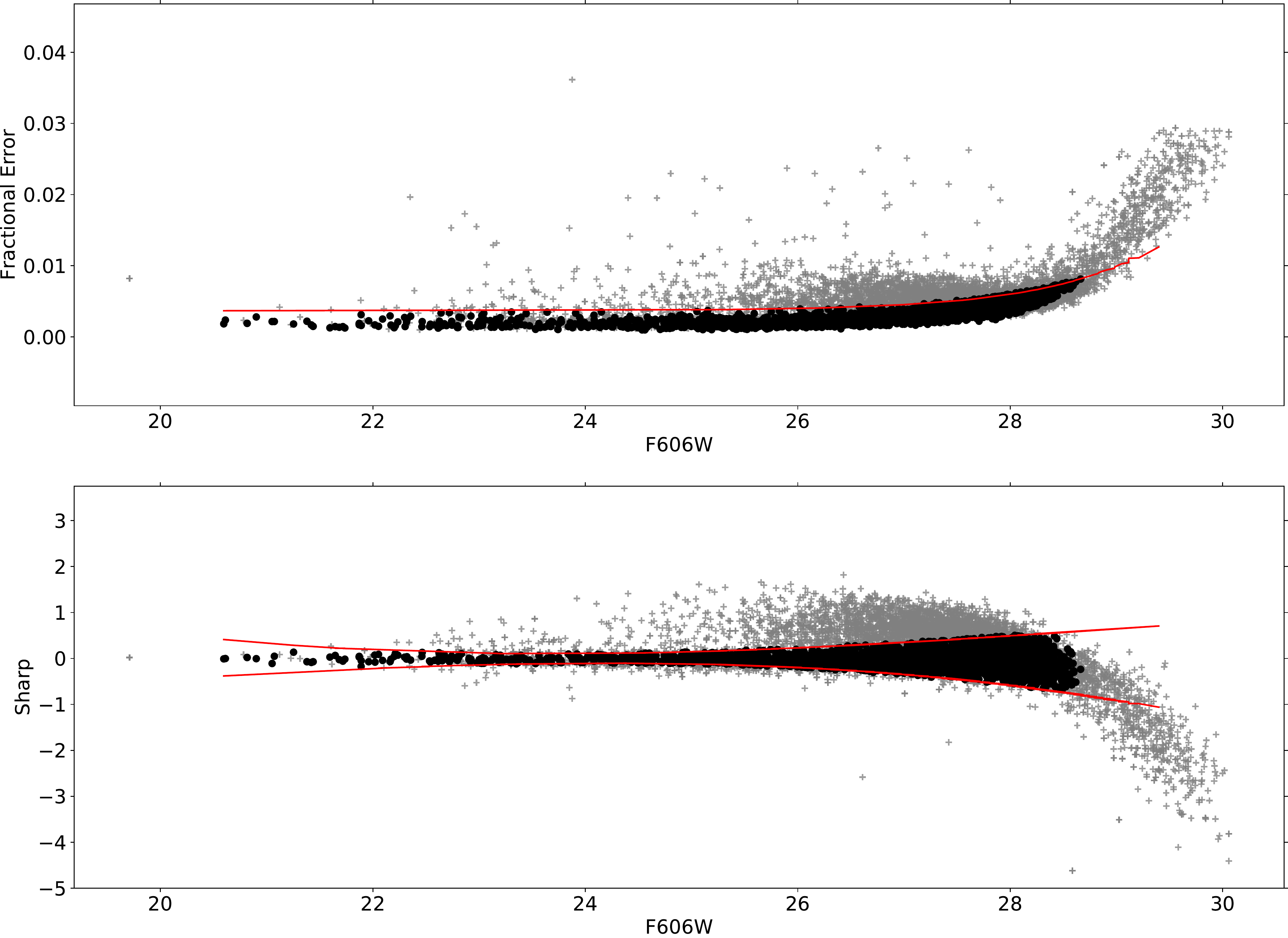}

\caption{The {distribution of the} DAOPHOT \textit{sharp} parameter {(lower panel)} and the fractional error {(upper panel)} versus the F606W magnitude {for sources detected in the ACS/WFC data.}  The red lines {in the lower panel enclose plus/minus}  three standard deviations from the mean {value for} the \textit{sharp} {obtained from the artificial star tests, estimated in magnitude bins of width 0.05~magnitudes. The red line in the upper panel indicates three standard deviations from the mean of the fractional error from the artificial star tests, calculated in the same magnitude bins.}  The sources plotted as black circles are those that pass all cleaning {criteria} in both of the F814W and F606W filters. The sources plotted as grey crosses {were}  rejected by the cleaning cuts - they are too dissimilar from the ideal PSF statistics, {as defined through the artificial star tests,}  to be identified as stars, {by one or more of the criteria, even if they pass the requirements shown in this plot.}}

\label{fig:cleaning}

\end{center}

\end{figure}

\begin{figure*}[ht]

\begin{center}

\includegraphics[width=\textwidth]{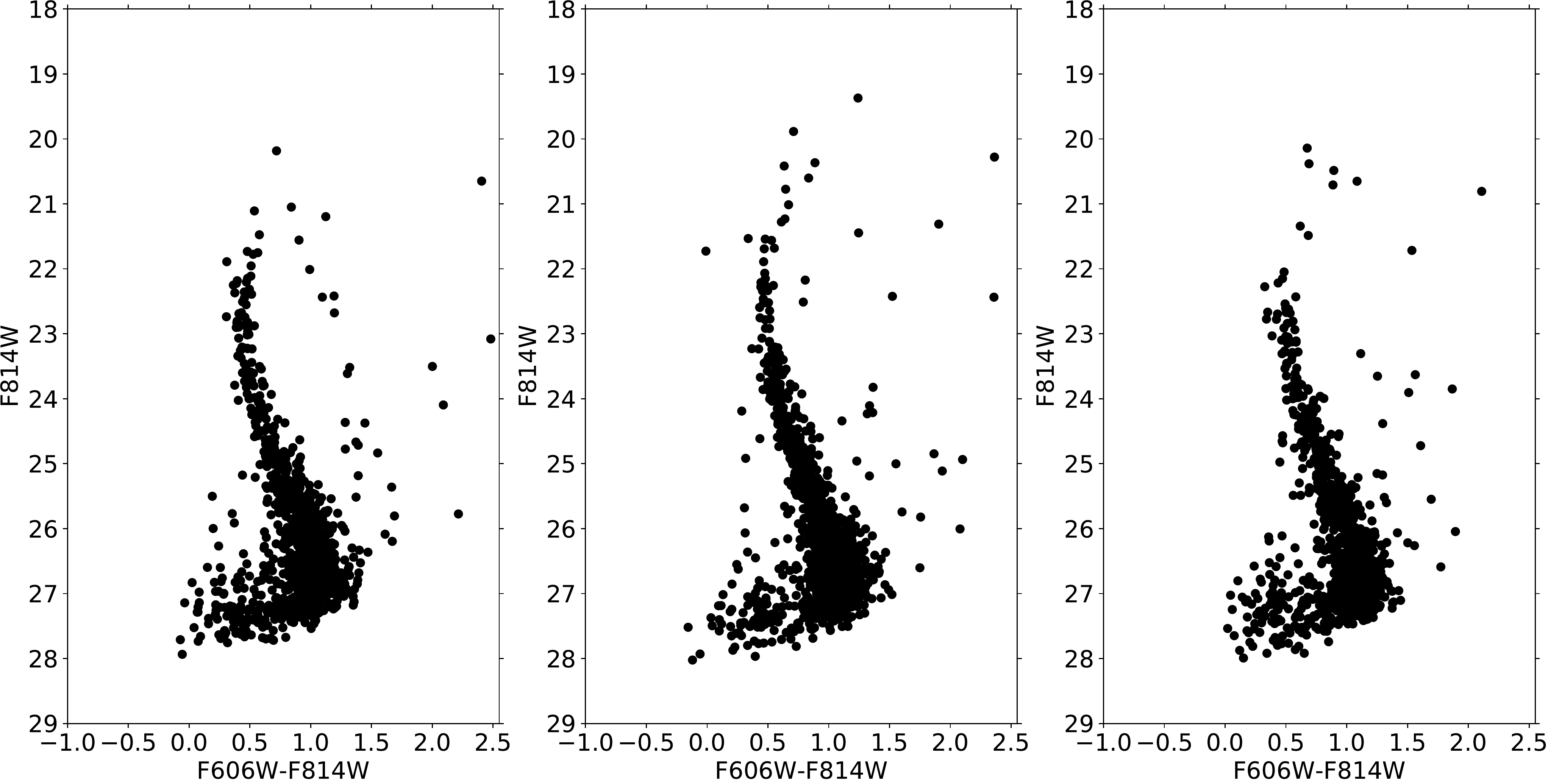}

\caption{The cleaned DAOPHOT photometric data for each of the three ACS/WFC pointings. As in Figure \ref{fig:epsf}, the number of sources retrieved per pointing is generally consistent with the surface density contours in Figure \ref{fig:footprint}.
Pointing one (left panel) has 927 sources.  Pointing three (middle panel) is closest to the center of \boo\ and has 1,167 sources. Pointing five (right panel) has 941 sources.}

\label{fig:allcmd}

\end{center}

\end{figure*}

{As noted, the artificial star tests also allowed for the determination of the completeness of the sample. First, stars with locations within problematic areas of the image, such as in masked regions or too close to the edges, were removed. Next, both the  input and retrieved artificial stars were binned into 0.05 magnitude bins,} spanning the input magnitude range. The ratio $N(\text{retrieved})/N(\text{input})$ for each magnitude bin then determined the completeness of the photometry, as shown in  Figure~\ref{fig:completeness}. {The data have a $50\%$ completeness limit of  27.4 for F814W and 28.2 for F606W (both on the Vega magnitude system) and this is the effective depth that we adopted for our further analysis. For comparison, the faint  limit reported in F814W for the previous, shallower HST ACS/WFC data \citep{gennaro} is 27.5 STMAG, corresponding to $\sim 26.2$ in the Vega magnitude system.} {We checked our retrieval process by cross-matching our final catalog with that of  \cite{brown}. We found a match, using a matching radius of $0.5\arcsec$,  for $92\%$ of our catalog brighter than STMAG of 26, and for $82\%$ of our catalog down to the faint limit given in \cite{gennaro}.}

\begin{figure*}[ht]

\begin{center}

\includegraphics[width=\textwidth]{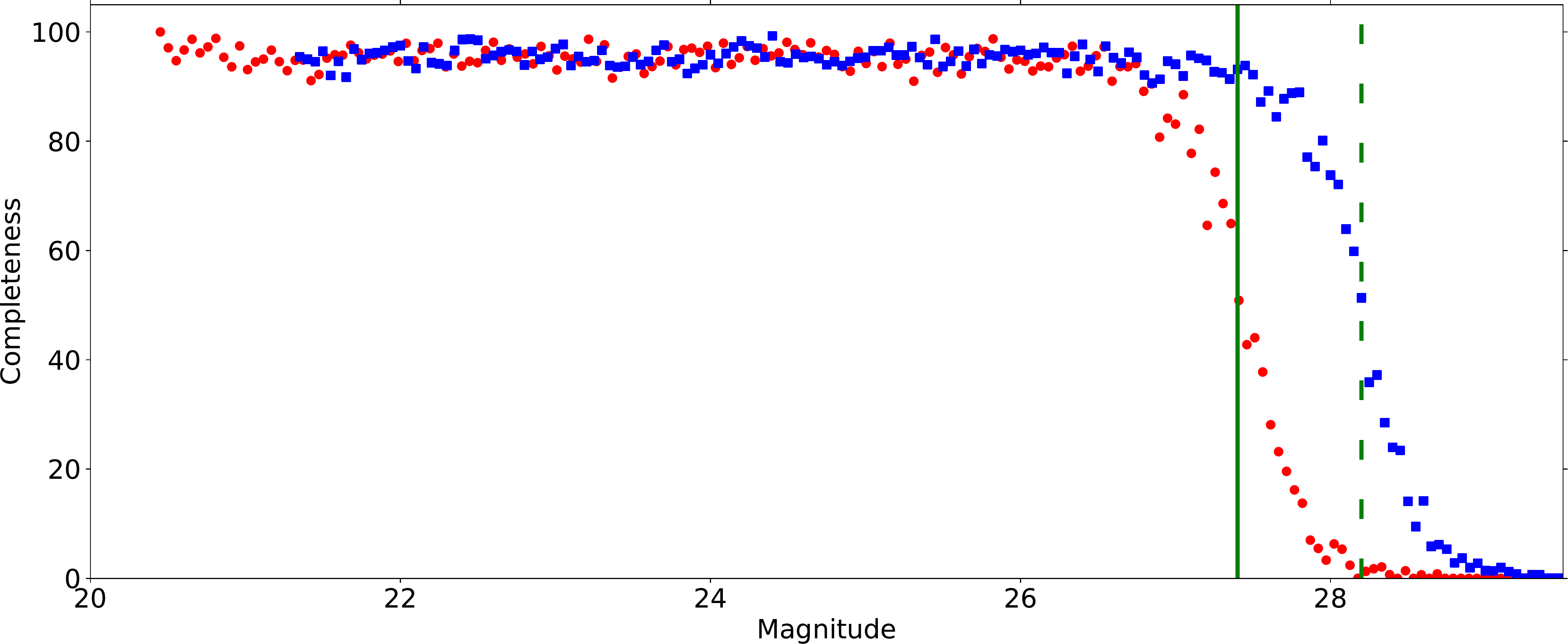}

\caption{Completeness {functions for the ACS/WFC observations, calculated from the artificial star tests as  $N_{retrieved}/N_{added}$ in magnitude bin of width $0.05$,} for F606W (blue squares) and F814W (red circles). The $50\%$ completeness magnitude for F606W is indicated {by the vertical} dashed green line at 28.2, and {that} for F814W {by the vertical}  solid green line at 27.4.}

\label{fig:completeness}

\end{center}

\end{figure*}

\subsection{Comparison between \textit{ePSF} and DAOPHOT} \label{sec:ePSFcomp}

We investigated the agreement between the \textit{ePSF} and DAOPHOT photometry to determine the optimal catalog and photometry pipeline for the remainder of our analysis. The color magnitude diagram resulting from each pipeline are shown side-by-side in Figure \ref{fig:daophotepsfcmd}. We matched the photometric catalog from each technique to compare the photometry, which required finding the shifts and scaling between the coordinate systems used in each analysis. The X and Y positions from \textit{ePSF} are in the ACS/WFC detector coordinate system, while the X and Y positions from DAOPHOT are in the HST V2\&V3 coordinate system \citep{acshandbook}. There is an offset and rotation between these coordinate systems, as well as a plate-scale difference due to the pixel size resampling done in \textsc{DrizzlePac}. The offset, rotation, and scale were solved for by aligning stars brighter than a magnitude of 24 in each catalog for each pointing, and then the cleaned PSF DAOPHOT catalogs were matched to the \textit{ePSF} catalogs. 
\begin{figure}[ht]

\begin{center}

\includegraphics[width=0.45\textwidth]{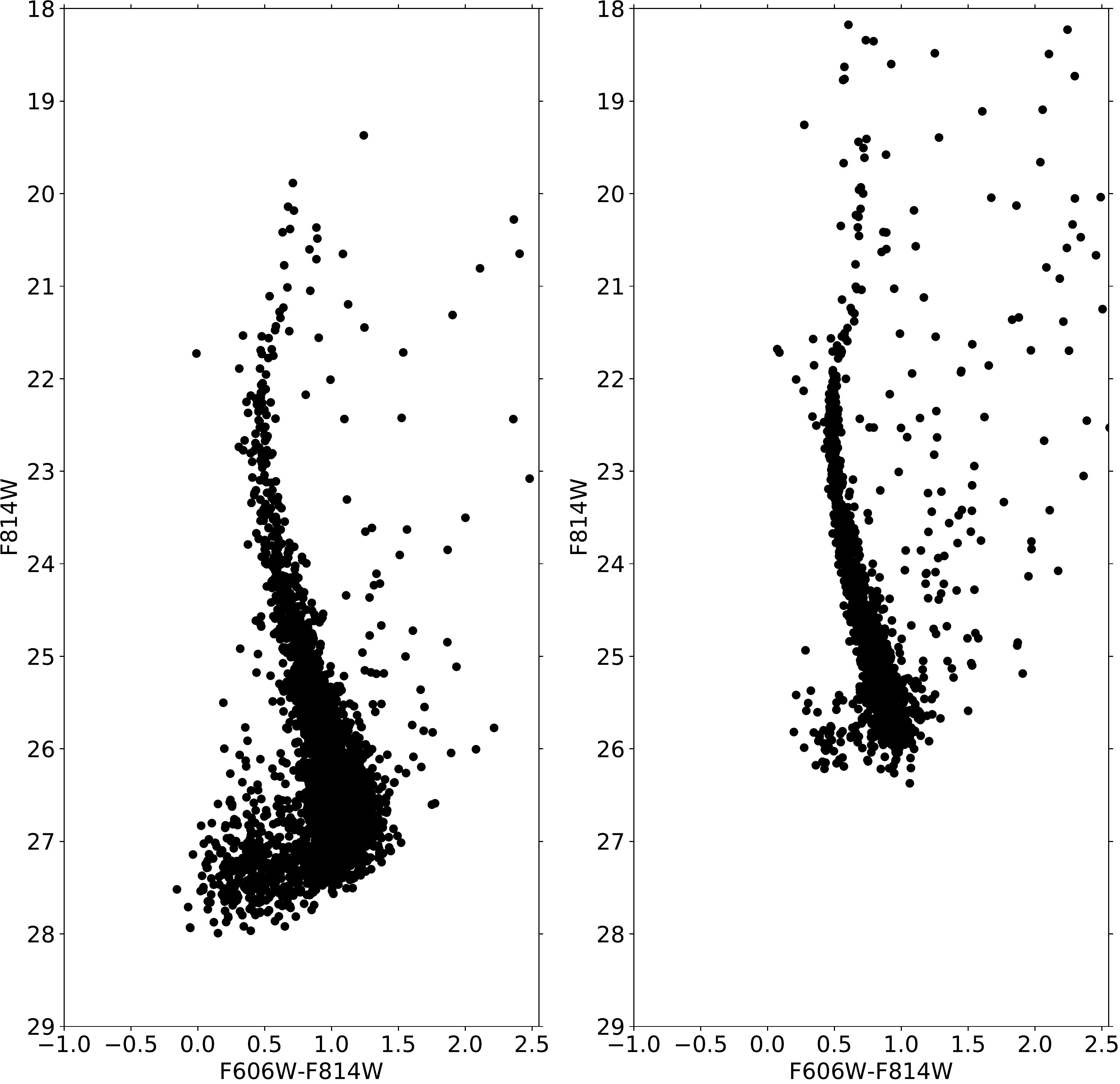}

\caption{DAOPHOT photometry (left panel) and \textit{ePSF} photometry (right panel), for all three pointings of the ACS/WFC observations. All sources consistent with being stellar are {included}. The metrics used to {determine} whether or not a detection is stellar {differs between the two photometric catalogs, as discussed in the text.}}

\label{fig:daophotepsfcmd}

\end{center}

\end{figure}

We determined that a rotation of two degrees, a shift of $(\alpha, \delta) \sim (+9.7\arcsec,-8.6\arcsec)$, and a scaling factor of $\frac{5}{4}$ were needed to match the \textit{ePSF} catalog to the DAOPHOT catalog. After the catalogues were matched, we found a mean magnitude offset of $\Delta m_{606} = -0.003$ and $\Delta m_{814} = -0.020$, where $\Delta m$ is the difference between the magnitudes from \textit{ePSF} and DAOPHOT, shown graphically in Figure~\ref{fig:deltamag}. {The small, but non-zero, mean magnitude offset likely stems from the differing inputs into our photometry pipelines (i.e. single exposures versus drizzle-combined exposures).} There is increased scatter at fainter magnitudes, as expected, {and a slight skew towards more negative residuals at faint magnitudes, especially for F814W}. The \textit{ePSF} photometry reaches to $\sim 26.3$ in F814W,  {brighter than the DAOPHOT $50\%$ completeness limit, but still  deeper than the previous dataset (from  GO-12549),} confirming that we achieved our goal of increased depth regardless of the choice of methodology of the photometry reduction. {Blue straggler stars have previously been identified in \boo\ as stars bluer and brighter than the dominant main-sequence turn-off \citep[e.g.][]{bootesdiscovery, okamoto}.} We note that the final \textit{ePSF} photometric {catalog} has a few more candidate blue straggler stars than {does the DAOPHOT catalog}. This difference is {likely due}  to the differing cleaning processes - some of the DAOPHOT {counterparts} to the \textit{ePSF} blue straggler candidates were {rejected on the basis of} their \textit{sharp} statistics.}
\begin{figure}[ht]

\begin{center}

\includegraphics[width=0.45\textwidth]{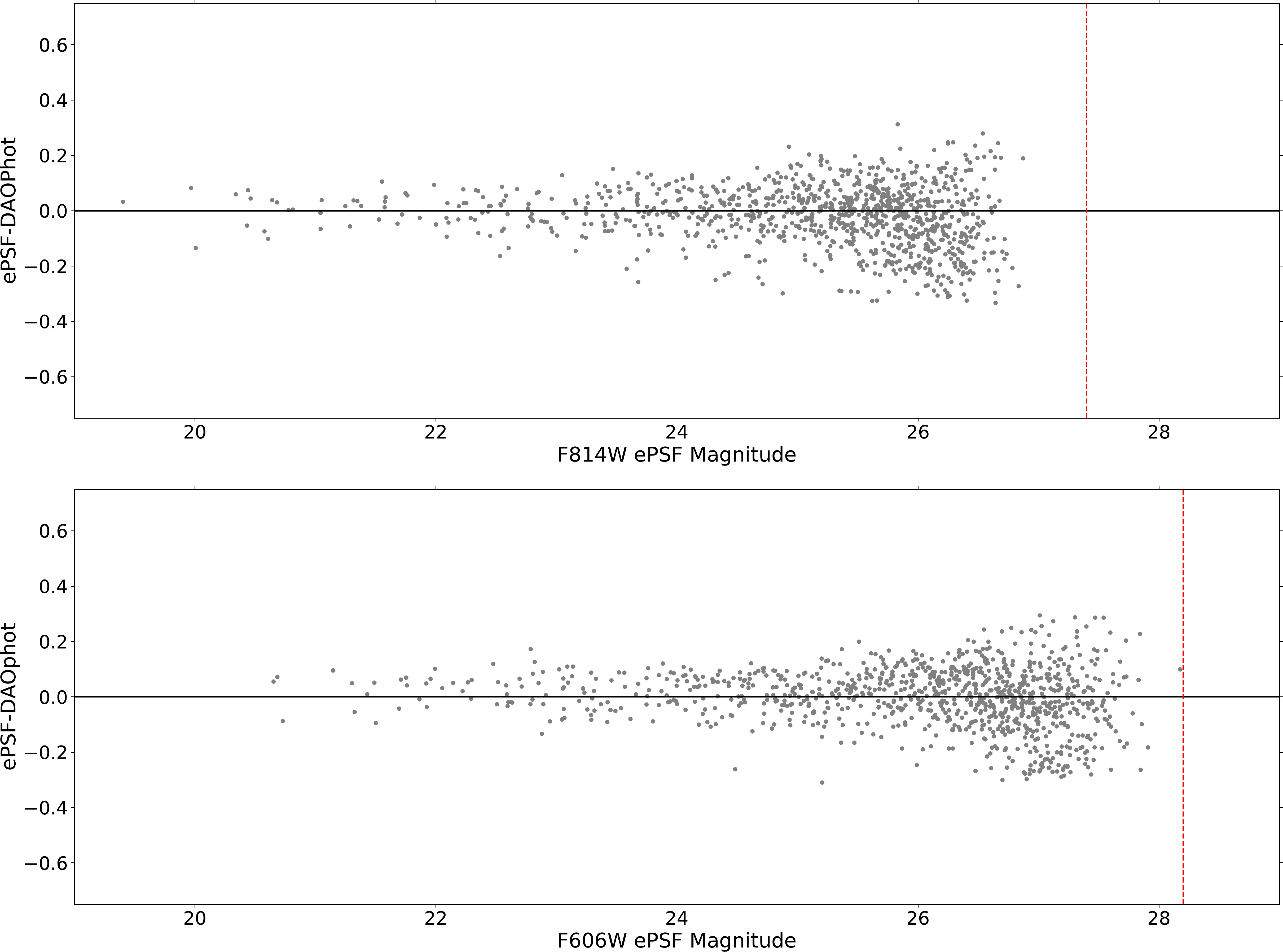}

\caption{The magnitude residuals between \textit{ePSF} and DAOPHOT {photometry} for {the combined} on-field ACS/WFC pointings, {as a function of apparent magnitude from} {\textit{ePSF}}, {for F814W (upper panel) and F606W (lower panel).  The DAOPHOT 50\% completeness level is shown as the vertical red dashed line, for context, and zero residual is indicated by the} solid black line. {There is excellent agreement at brighter magnitudes, with an increase in the residuals} at fainter magnitudes, as expected.}

\label{fig:deltamag}

\end{center}

\end{figure}

The observed residuals {depend on location on the image,  which is plausibly related to  the different levels of flexibility  in the PSF fitting routines with}  \textit{ePSF} and DAOPHOT.  The two-dimensional map of  the residuals in  magnitudes  between the two approaches is presented in Figure~\ref{fig:spatial_var}, where the non-linear variation across the ACS/WFC CCD chips is evident. This {map was produced} by  fitting a second-order polynomial surface to the three-dimensional data of the X and Y positions of each match between the \textit{ePSF} and DAOPHOT catalogues, and the residuals between the photometry. {This  spatial variation could be} due to three factors: first, the  PSF within \textit{ePSF} is an empirical model derived from {the data,  whereas the PSF within} DAOPHOT is an analytic Penny function {which was fit to the data}. Second, \textit{ePSF} creates 45 empirical models across each ACS/WFC CCD chip and interpolates amongst the four nearest grid points for any arbitrary point in the ACS/WFC exposure, {while within  DAOPHOT, the PSF has fixed functional form and} can vary  across the chip up to quadratic order {in the coordinates \citep{daophot91}}.  Third, the \textit{ePSF} library is created for each calibrated ACS/WFC filter, {while} DAOPHOT uses the same base analytic function {(a Penny function for the present  discussed)} and fits it to the data for each filter.  The {somewhat lower flexibility of DAOPHOT relative to \textit{ePSF} results in a reduced ability to capture fully the spatial variation of the true PSF across the CCD chip (caused by variations in chip thickness, \citealp{anderson,krist}). {The residuals can be either positive or negative for F606W, but are all negative for F814W. The all-negative values in the fit to the magnitude residuals for F814W, and the larger negative values in F606W, both come from the relatively higher number of detections with negative residuals at faint magnitudes, as seen in Figure \ref{fig:deltamag}.} {The amplitude of the residuals has a maximum value of $\sim 5\%$ at most, and  reaches this extreme only} in the very corners {of the chip,} for each filter, where sources were less likely to benefit from the dithering pattern. { Thus while DAOPHOT may not perfectly model}  the variation of the PSF across the CCD chip, the average residual between the photometry derivations is low (less than 0.03 magnitudes), {and DAOPHOT photometry reaches a fainter limit than does \textit{ePSF}.   We therefore adopt the DAOPHOT photometry catalog for the analysis of the low-mass stellar IMF (in paper II, Filion et al.~in prep.)}
\begin{figure*}[ht]

\begin{center}

\includegraphics[width=\textwidth]{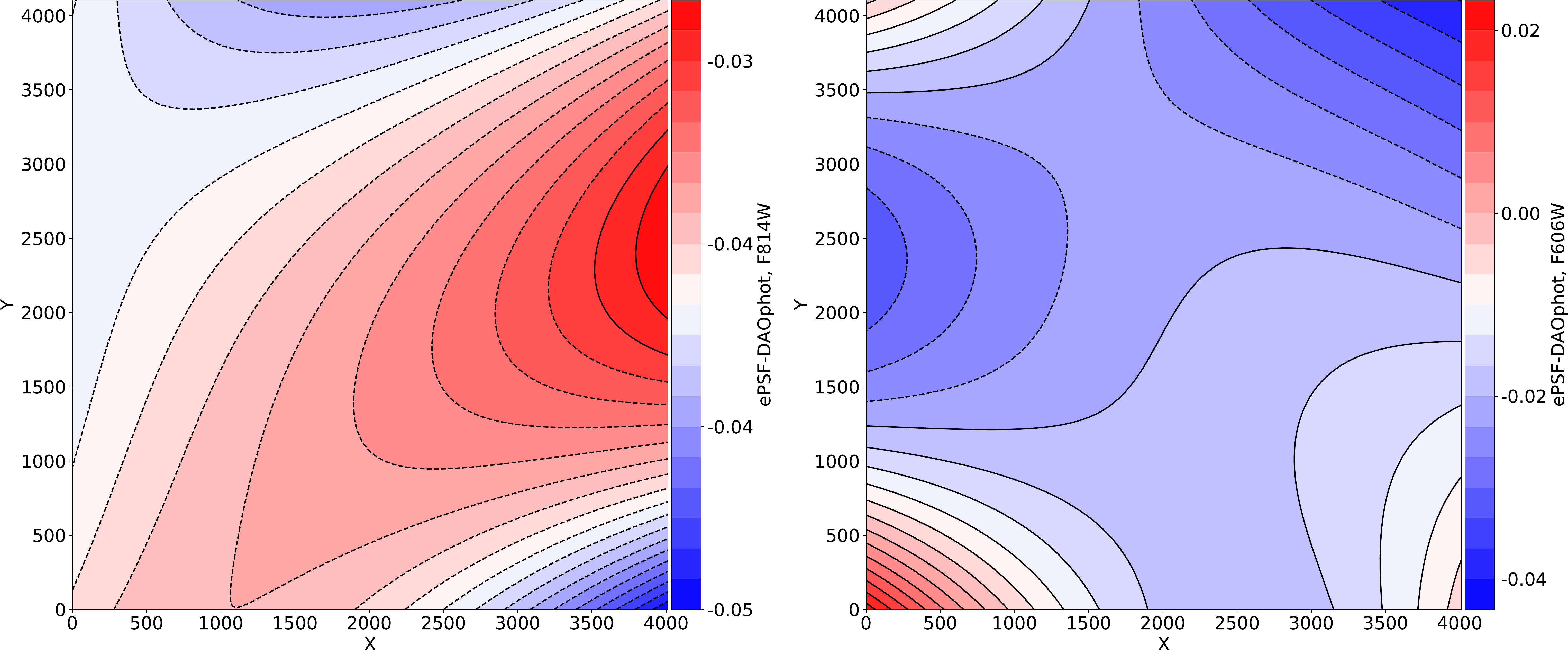}

\caption{{Two-dimensional maps of the  residuals, in magnitudes, between the} \textit{ePSF} and DAOPHOT photometry for F814W (left panel) and F606W (right panel), for the ACS/WFC observations. {The maximum amplitude of the residuals, corresponding to $\sim 5\%$, occurs towards the edges of the images. Typical}  residuals are $\simlt 3\%$.}

\label{fig:spatial_var}

\end{center}

\end{figure*}
\subsection{Off-Field Observations} \label{sec:offfield_phot}

The off-field observations taken under our program (GO-15317) consist of {one continuous exposure} in each filter. As before, we used the bias corrected, flat-fielded and  CTE-corrected data. We ran these off-field exposures through \textit{AstroDrizzle} to correct for distortion \citep{vera} and to produce output images of the same {dimensions as the data for the on-field} pointings. We then ran LACosmic \citep{dokkum,astroscrappy} to remove as many cosmic rays as possible, {setting the threshold detection limit for cosmic rays to be}  four standard deviations above the background. {A significant number of cosmic rays remained, however.}  DAOPHOT was then run on {these cleaned} exposures, following the same procedure as for the on-field data. {Unfortunately, after inspection of  the sources that were retrieved, we determined} that the exposures {remained too} degraded by cosmic rays to be of use {for our purposes  and we proceeded with the analysis of data for an alternative   off-field that we identified in the HST archive.}

We searched the MAST archive for fields of similar Galactic latitude and longitude to {those of Boo~I (given in Table~\ref{tab:boo-param}), with the  further restrictions that the data have been taken in the  same filters and with the same aperture (WFCENTER) as the on-field observations, and have similar exposure time to the original off-field data.} We avoided fields with coordinates close to neighboring satellite galaxies such as Boo~II or other known stellar over-densities, as the goal of the off-field data is to gain an understanding of the average Galactic contributions to { the star counts in the Boo~I fields}. The observations that we selected are from GO-13393, {in the line-of-sight with Galactic coordinates ($\ell,\, b {\rm ) = (331^\circ.6, +46^\circ.7)}$, hereafter denoted by} Off Field 2, and consist of four images per filter {(see Table~\ref{tab:obs})}.

We combined the individual images with \textit{AstroDrizzle} using the same parameters as the on-field ACS/WFC exposures. Photometry was done in \textsc{DAOPHOT}, again following the same procedure and conversions as for the on-field ACS/WFC exposures. Approximately 80 stars were retrieved in this field, {the color-magnitude diagram for which is} presented in Figure~\ref{fig:contcomp} alongside the {color-magnitude diagram we obtained for \boo\  and predictions from the Besan{\c c}on model of the Milky Way Galaxy at the appropriate coordinates}.  {Note the higher number of Milky Way stars predicted for  Off Field 2, reflecting its  lower Galactic} latitude.  This field also has higher extinction than {the line-of-sight towards \boo\ and}  following the procedure from Section~\ref{sec:dust} we obtained $(A_{606}, A_{814}) = (0.148, 0.096)$.

\section{Photometry: WFC3/UVIS Data}

{The UVIS data were obtained in parallel mode, during the primary observations of fields in \boo\ with ACS/WFC. WFC3/UVIS has a lower throughput than ACS/WFC \citep{dressel}, and hence our observations with UVIS are not as deep our observations with ACS/WFC. These observations form the basis of our analysis of the star-formation history of \boo\ (Paper~III), where photometric precision for stars on the main sequence turn-off is crucial. We produced photometry for these fields using \textit{ePSF}, as the high precision of \textit{ePSF} photometry best fulfils these science goals.}

{The \textit{ePSF} UVIS photometry procedure {followed that for the ACS/WFC data,}  differing only in the values used for {the photometric conversions}. We adopted the appropriate Vega magnitude zero points for UVIS observations, and {took} the encircled energies from the UVIS website\footnote{\url{https://www.stsci.edu/hst/instrumentation/wfc3/data-analysis/photometric-calibration/uvis-encircled-energy}}. The final \textit{ePSF} color-magnitude diagram for each pointing is shown in Figure~\ref{fig:uvis_cmd}, {where it may be noted that the  limiting magnitude is not quite as faint as for the ACS/WFC photometry, as anticipated due to their different sensitivities.} As before, only sources present in at least three exposures per filter are presented. The smaller pixel scale of UVIS compared to ACS/WFC has allowed for more precise photometry, which can be seen in the tighter main sequence and turnoff region of each pointing of Figure \ref{fig:uvis_cmd}.}

\begin{figure*}[ht]

\begin{center}

\includegraphics[width=\textwidth]{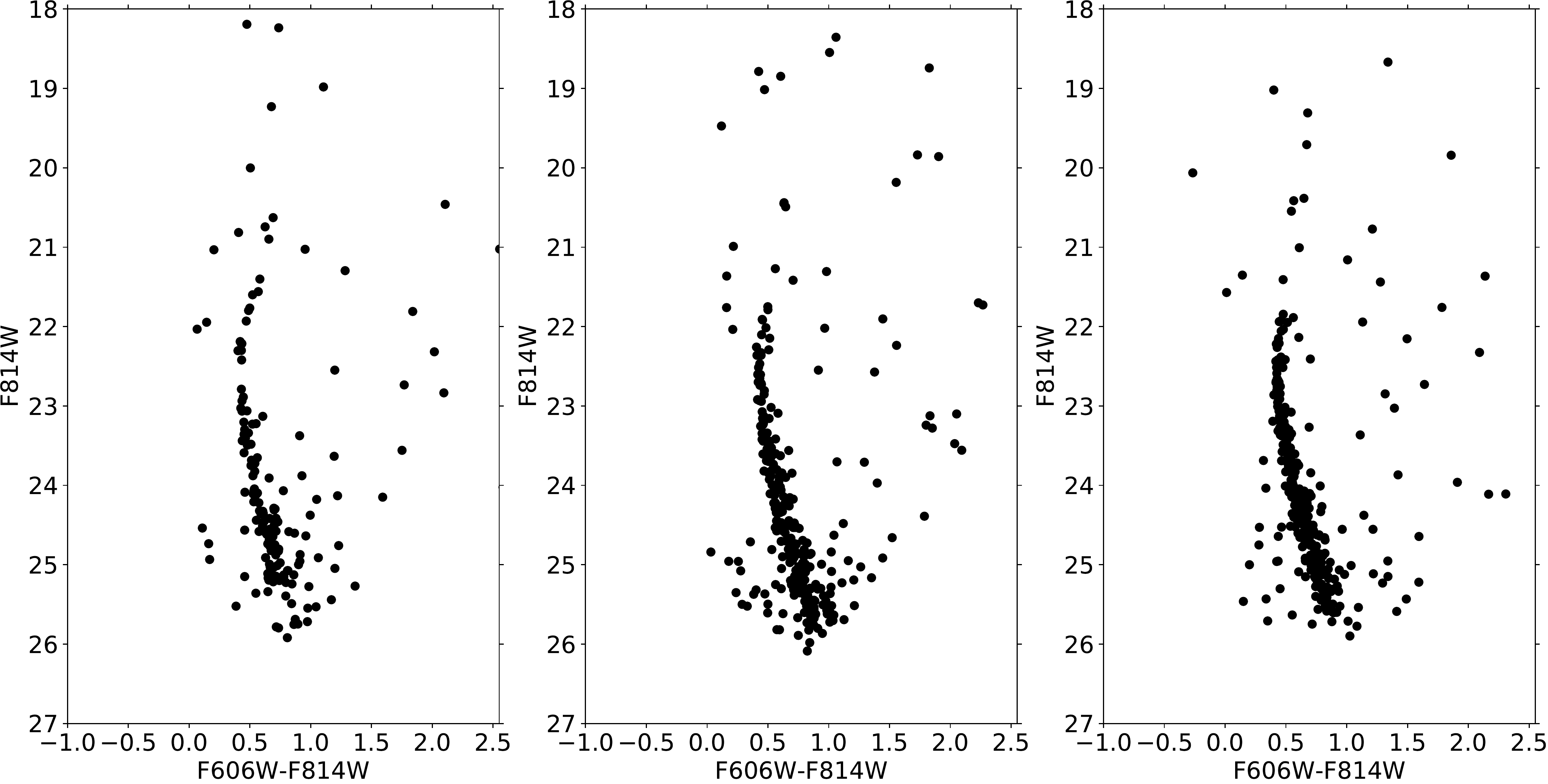}

\caption{{The} color-magnitude diagram for each field of the UVIS data produced with \textit{ePSF}. Only sources present in {at least three} exposures per filter are shown. Pointing A with 171 sources is leftmost, B with 309 sources is central, and C with 339 sources is rightmost. }

\label{fig:uvis_cmd}

\end{center}

\end{figure*}


\section{Membership Selection} \label{sec:CMD}
One of the more difficult aspects of photometric-only studies of {the resolved stellar population of} external galaxies is {discerning which detections are likely members of the galaxy, and which are foreground/background contamination}. To this end, {the population(s) of the galaxy are usually represented by theoretical isochrones and stars rejected or accepted on the basis of proximity to the isochrone in the color-magnitude plane.  Previous}  studies \citep[e.g.][]{wyse, bootesdiscovery} simply defined a polygon encompassing a single {isochrone, with properties matching those of the dominant population,} and define all stars falling inside of the polygon as members, discarding those falling outside of the selected region. {We developed a}  statistical method {that builds on this approach and assigns a weight to each star} based on {its distance in color from a pair of isochrones, selected to bracket the known properties of \boo, and  that takes photometric errors into account. A high weight indicates a location on the color-magnitude diagram that is consistent with membership of \boo\ and, conversely, a low weight indicates a non-member.}

 \cite{roderick2015} devised a weighting scheme using Gaussian distributions in color space, centered on an isochrone representative of the mean population of {the galaxy under study. We adopted a similar weighting scheme, but extended  the technique through the use of two isochrones.  As noted earlier, there is a significant metallicity spread in \boo\ and this manifests  itself through a broadening in color of the lower main sequence, below $\sim 0.5 \msun$, where sensitivity to metallicity is maintained, even at the low values relevant for \boo.}   The goal of this work was to reach $\le 0.3 \msun$  and thus it was especially important to devise a scheme that incorporated both the effect of the metallicity spread and the increasing photometric error at fainter magnitudes.

 { We selected} isochrones from the Dartmouth  Stellar Evolution Database \citep{dartmouth},    as they {allow fine-sampling of  the low-mass regime of  interest, plus}  they provide alpha-enhanced isochrones in {the appropriate photometric bands for both ACS/WFC and WFC3/UVIS}. We defined the most metal-poor and the most metal-rich isochrones to use by  taking into account the trend of {decreasing alpha-abundance enhancement with increasing iron abundances seen in \boo\ member stars, noted above}  in Section \ref{sec:boo}. We adopted ${\rm [Fe/H]} = -3.5$, ${\rm [\alpha/Fe]=0.5}$ for the lowest metallicity, and ${\rm [Fe/H]} = -1.8$, ${\rm [\alpha/Fe]=0.1}$ for the highest metallicity. We {obtained the appropriate metallicities (${\rm [M/H]}$) of the isochrones from} the ${\rm [Fe/H]}$ and ${\rm [\alpha/Fe]}$ values using the   {transformation}  determined by \cite{salaris} and refined by \cite{salarisbook}:
$${\rm {\rm [M/H]} = {\rm [Fe/H]} + \log(0.694 \times 10^{[\alpha/Fe]} + 0.306)}$$  This gave {values of} ${\rm [M/H]} = -3.1$ and ${\rm [M/H]} = -1.7$ for the lowest and highest metallicity isochrones to be used in defining the {properties of \boo\ stars in color-magnitude space}. However, the most metal-poor isochrone {that covers the required low-mass range that is} provided by \cite{dartmouth} has ${\rm [M/H]} = -2.5$ (isochrones of ${\rm [M/H]} = -2.688$ and ${\rm [M/H]} = -3.188$ are available, but {lack entries for} masses below $\sim 0.65 \msun$). {Happily, the sensitivity of the color along the lower main sequence to metallicity is predicted to saturate below $\sim -2.5$~dex,} with minimal difference between {isochrones with} ${\rm [M/H]} = -2.5$ and ${\rm [M/H]}=-3.0$ {(A.~Dotter, priv.~comm.)}  {In light of this, plus the advantages of this set of isochrones noted above, we adopted ${\rm [M/H]}=-2.5$ as the lowest metallicity isochrone, even though this is}   $\sim 0.5$ dex higher than the lowest {published metallicity for a radial-velocity member of  \boo\ \citep{norris_2010a}}.  We {adopted a fixed}  age of 13~Gyr  and shifted the isochrones to {apparent magnitudes and colors appropriate for an assumed distance of $62$~kpc and reddening and extinction from dust, as}  discussed in Section \ref{sec:photandastACS}. {A plot of these isochrones overlaid on the DAOPHOT ACS/WFC color-magnitude diagram for the \boo\ fields is presented in Figure~\ref{fig:isochronecmd}.}
\begin{figure}
\begin{center}

\includegraphics[width=0.45\textwidth]{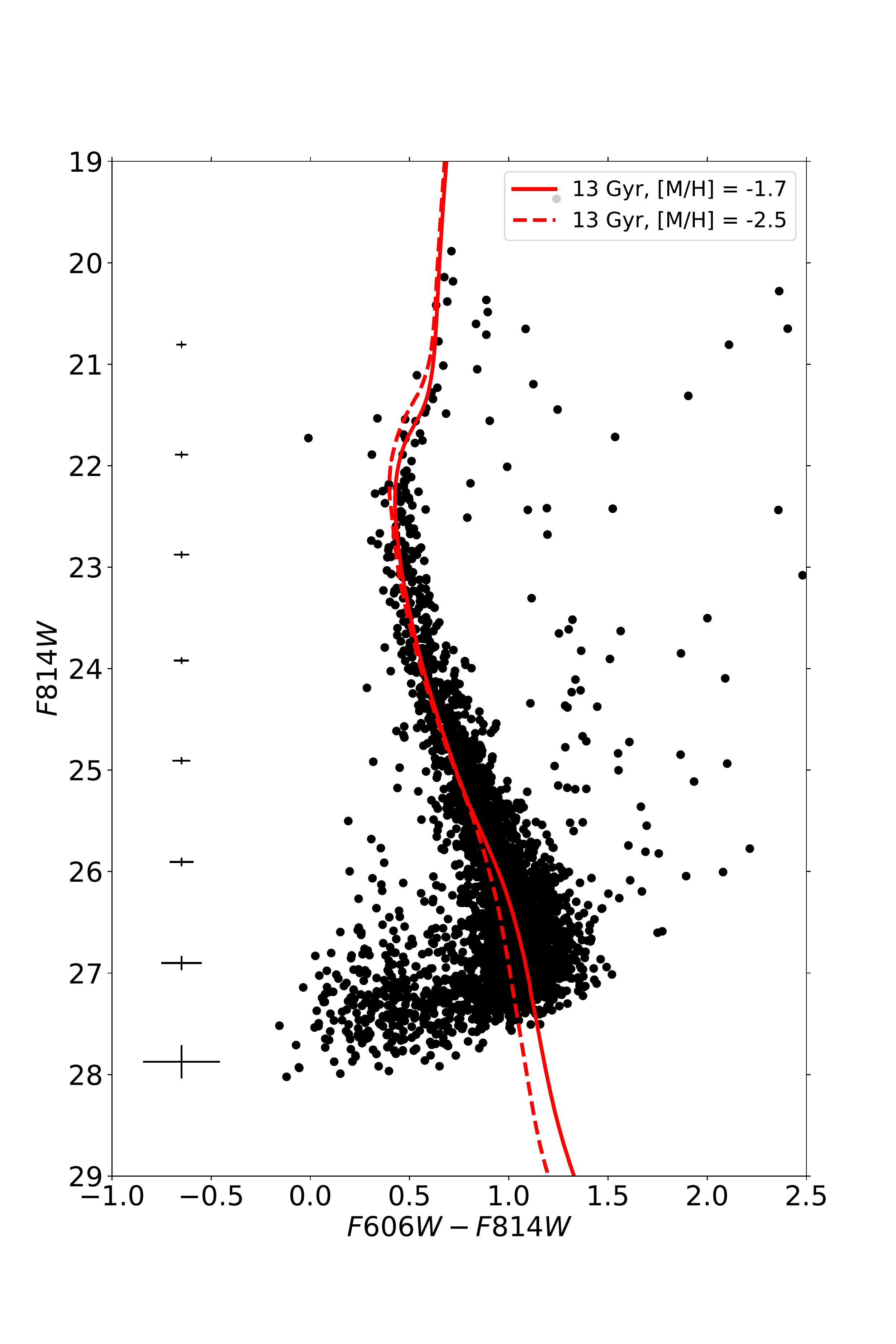}

\caption{The cleaned DAOPHOT color-magnitude diagram for all three on-field ACS/WFC paintings combined. The solid red curve corresponds to a 13~Gyr, [M/H] = -1.7 isochrone, and the dashed curve a 13~Gyr, [M/H] = -2.5 isochrone, {chosen to represent the observed range in metallicity in \boo. Both isochrones have been corrected for dust and shifted to  apparent magnitudes appropriate for an  assumed distance of 62~kpc.}  {The photometric error at fiducial magnitude levels, estimated following the discussion} in Section~\ref{sec:DAOphot}, are {indicated on the figure, at left. Note the divergence of  the isochrones  at lower masses, due to the} increased sensitivity of the opacity to metallicity in this regime.}

\label{fig:isochronecmd}

\end{center}

\end{figure}

 We investigated the {identification}  of non-\boo\ contaminants by {first defining} Gaussian distributions {in color,} centered on each {of the two} isochrones described above, {and with mean and dispersion varying with the apparent magnitude along the isochrone.}  The mean of each distribution equaled the color  of the dust-adjusted isochrone at that magnitude, while we set the dispersion to be three times the mean photometric error of the data within a bin of  width 0.01~magnitudes, centered at that magnitude.   {We then defined the membership weight, $w_i$ for a given star of color $c_i$, F814W apparent magnitude $I_i$, with corresponding mean color error $\sigma_i$ as:} 
$$w_i = \frac{1}{2}  \big[\exp{\frac{-(c_i - \mu_{1})^2}{2(3 \sigma_i)^2}} + \exp{\frac{-(c_i- \mu_{2})^2}{2(3 \sigma_i)^2}}\big],$$ where $\mu_1$ is the color of the ${\rm [M/H] = -1.7}$ isochrone at $I_i$, and $\mu_2$ is the color of the ${\rm [M/H] = -2.5}$ isochrone at $I_i$. {Thus a star lying close to either of the isochrones will be assigned a weight close to unity while stars that lie far from either isochrone will be assigned a low weight. The most straightforward application is to identify non-members.}

   This technique {can be thought of as defining membership by drawing a polygon around an isochrone, but} where the width of the polygon {at any apparent magnitude is set  by the choice of threshold value for the weight, below which a star is rejected as a non-member.  The fact that photometric errors are incorporated into the calculation of the weight provides for a widening at fainter magnitudes even for fixed threshold in the weight. The physical widening of the lower main sequence due to the sensitivity of the color of low-mass stars to metallicity is captured through the use of the two isochrones that represent the range in metallicity of stars in \boo. This technique does not account for the shape of the metallicity distribution, however, which results in some stars lying between the isochrones receiving a weight less than unity.}
   
   {The color-magnitude diagram color-coded by the weights calculated in this manner is shown in Figure~\ref{fig:weightedcmd}.} {We cross-matched our {catalog with spectroscopic studies of \boo\ in the literature, to test} our technique. We found four stars with matches, two {radial-velocity \boo\ member stars from \citet{koposov} and one radial-velocity member plus} one non-member from \citet{martin_met}. Our technique assigned {a weight above}  $0.7$ to each of the {radial-velocity members}  and a {very low} weight ($\sim 0$) to the non-member. {These stars with spectroscopic information are indicated by special symbols  in Figure~\ref{fig:weightedcmd}. {It is clear from the Figure that adopting a threshold  weight of $\sim 0.5$ would separate stars occupying the locus of the  main sequence and turn-off region from those that are more dispersed across the color-magnitude diagram.}

\begin{figure}[ht]

\begin{center}

\includegraphics[width=0.5\textwidth]{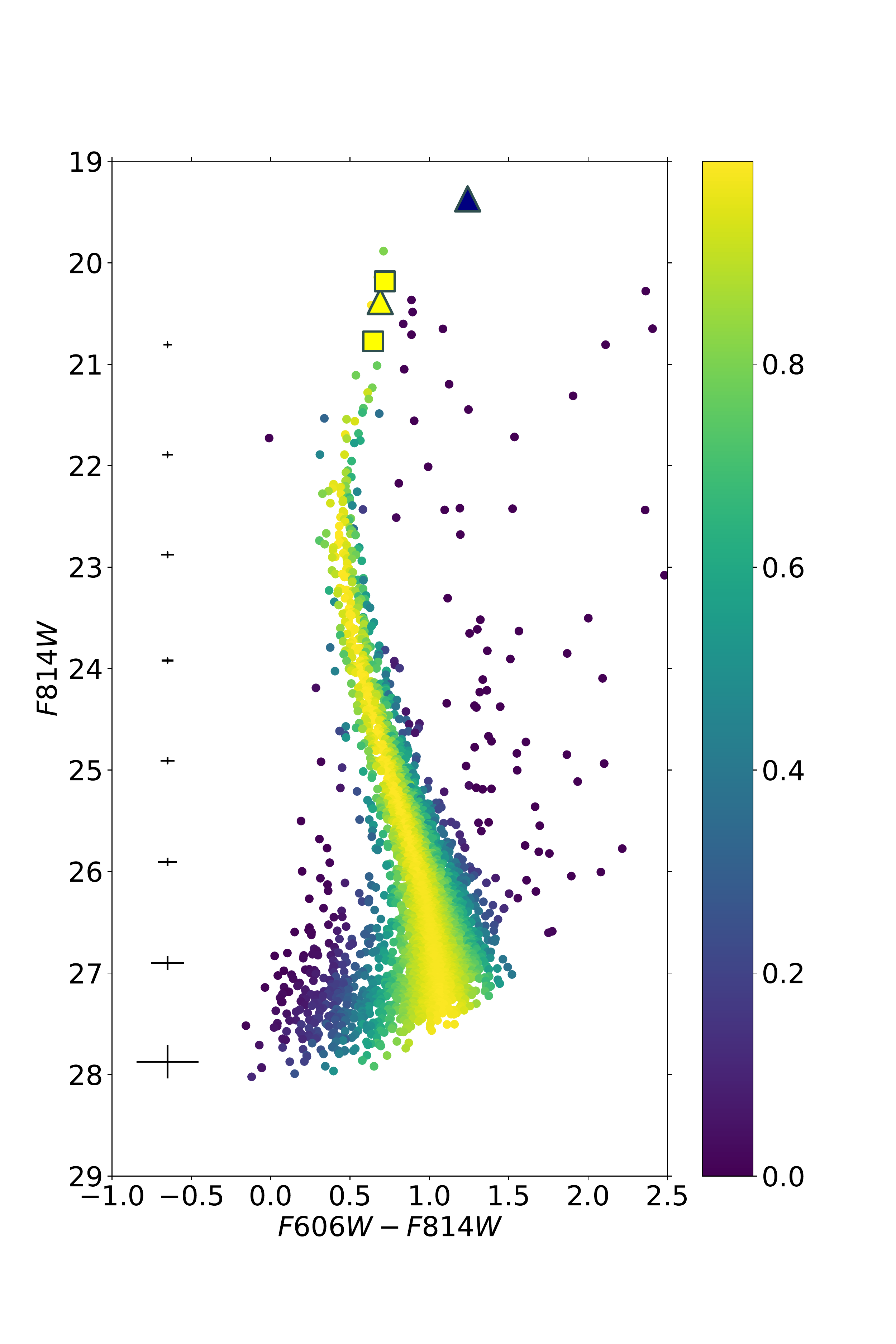}

\caption{{The cleaned DAOPHOT}   color-magnitude diagram of the on-field ACS/WFC data, {as shown in Figure~\ref{fig:cmd}, now} colored by weight, {as indicated by the color bar at right. The high-weight stars (yellow) fall along the main sequence and turn-off region, while low-weight stars (blue) lie either to faint magnitudes with blue colors or redward of the main sequence. The yellow (high-weight) square  symbols indicate radial-velocity member stars from  \citet{koposov} and the yellow triangle indicates a radial-velocity member from \citet{martin_met}. The navy (low-weight) triangle indicates a field star from \citet{martin_met}.}}

\label{fig:weightedcmd}

\end{center}

\end{figure}

       Thus far we have ignored the presence of binary systems, which would lie to the red of the single-star main sequence, and indeed Figure~\ref{fig:weightedcmd} shows a band of low-weight sources occupying that locus.  We therefore defined `equal-mass binary isochrones' by shifting the two isochrones brighter in apparent magnitude, subtracting 0.75~mag. We then re-calculated the weights {based on position relative to  these shifted isochrones. This procedure returned $108$ stellar sources, or $3.4\%$ of our full sample, which received a weight below 0.5 using the single-star isochrones but received a weight above this value using the `equal-mass binary' isochrones. These sources are identified in         Figure~\ref{fig:cmd} and    could be unresolved binary systems.}
\begin{figure}[ht]

\begin{center}

\includegraphics[width=0.5\textwidth]{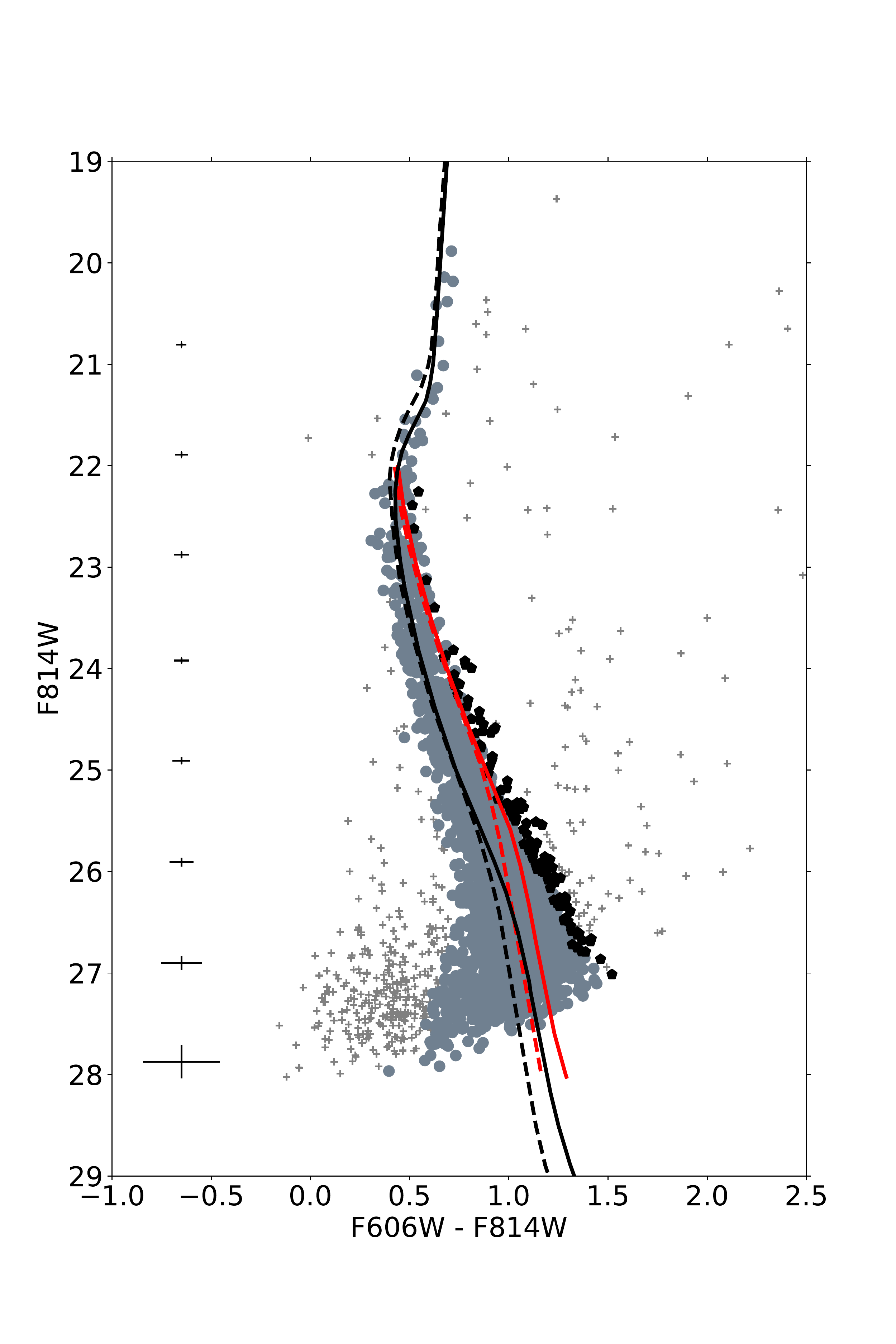}

\caption{The cleaned, DAOPHOT color-magnitude diagram as in Figure~\ref{fig:cmd} and Figure~\ref{fig:weightedcmd}, now with different symbols to indicate weight and possible binarity. {The solid black line indicates a 13~Gyr, $\rm[M/H] = -1.7$ isochrone, and the solid red line is this isochrone shifted brighter by 0.75 magnitudes, to mimic an  equal-mass binary sequence. The dotted black line is a 13~Gyr, $\rm[M/H] = -2.5$ isochrone, and again the dotted red line is this isochrone shifted 0.75 magnitudes brighter. The grey filled circles represent  stars that have weight greater than 0.5 based on distance from the unshifted isochrones, while  the filled black pentagons, which occupy the locus of possible unresolved binaries,  have weight greater than 0.5 based on the shifted isochrones, but below 0.5 on the unshifted isochrones. The grey crosses represent all other sources consistent with being stellar.}}

\label{fig:cmd}
\end{center}

\end{figure}

         Assuming the $28\%$ binary fraction found for \boo\  in \cite{gennaro}, this translates to roughly $12\%$ of all binaries being rejected by cutting on a weight of $0.5$, if every one of the selected sources is an unresolved binary member of \boo. Though relative to our total sample, the number of potentially rejected binaries is small, the likely possibility that this technique could reject true members of \boo\ depending on the selected weight value motivates the more {sophisticated}  Bayesian approach we adopt in {Paper~II. }

         {As noted earlier, blue straggler stars should also be present within \boo,  but since these are plausibly evolved from  close-binary systems they also will not be properly treated in any weighting scheme that was based on  single-star stellar evolution models.  The blue sources near the main sequence turn-off in the colour magnitude diagram may well be members of \boo, but these  candidate blue stragglers must be treated independently.}

\subsection{Non-Member Contamination}\label{sec:contdisc}
           {The two main sources of possible contamination in the lines-of-sight towards \boo\ are  background galaxies and  foreground stars}. The background galaxies should have been largely removed {during the artificial-star tests and associated cleaning of the DAOPHOT photometry routines, due to their non-stellar PSF. However, faint, blue high-redshift galaxies are expected to be barely resolved even in HST images \citep{bedin} and will not be removed. Similarly, non-member stars will be included in the DAOPHOT catalog.}  We {estimated the importance of these sources of contamination through  both the predictions of theoretical models of {the distribution of stars in the Milky Way and the analysis of  photometry for {an offset field with Galactic coordinates as well-matched as possible to those} of \boo.}

\subsubsection{Contamination by Milky Way Stars}

{We used the Besan{\c c}on model \citep{besancon} to make predictions for the contribution of stars in the Milky Way along the lines-of-sight of this study.} We created {simulated star counts for  Off Field 2 and the \boo\ fields.} The Besan{\c c}on model {assumes smooth density laws for each of the main stellar components, namely stellar halo, bulge, thick disk and thin disk.}  \boo\ {lies in the `Field of Streams' \citep{FoS} and several streams are close in projection and} may contribute additional contamination {beyond that predicted by the model.}  The Besan{\c c}on model does not provide output in HST photometric bands {and} we selected the Johnsons-Cousins {(V, I)} system for the output. {We then compared the predicted distribution of stars in the  color-magnitude diagram to an isochrone representative of \boo\, in this photometric system, rather than attempt to convert between the magnitude systems}.

{At these faint magnitudes and intermediate latitudes, the expected contribution from the Milky Way should be predominantly lower main-sequence stars and white dwarfs from the thick disk and halo.}  The {predicted distributions} of the Milky Way stars in color-magnitude space {are} shown in the {upper}  panels of Figure~\ref{fig:contcomp}, and  {indeed the stars are mostly low-mass dwarfs from the halo and thick disk and the majority falls significantly}  redward of the isochrone representative of {the mean population in} \boo.  The {predicted distribution shown in the upper  righthand} panel of Figure~\ref{fig:contcomp}, in the line-of-sight towards \boo, {contains}  fewer than $\sim 30$ Milky Way stars {which}  have colors and magnitudes overlapping with \boo\ member stars. {The remaining Milky Way contaminant stars, also $\sim 30$ in number,  would be  removed through an application of the membership weighting scheme described above.}{ Foreground stars should have a higher proper motion than do 
members of \boo\ and the cross-match with the earlier data from GO-12549 can be used to identify non-members, even those  similar in color and magnitude to \boo\ members. The time baseline from these first-epoch observations to our observations is 7~years, implying an expected shift in position for a star at distance $d$, with transverse velocity $v_t$, of amplitude $\sim 0.03\arcsec {\rm (} v_t/100{\rm km/s)}{\rm (5 kpc}/d)$. Remembering that the drizzled images have a scale of $0.04\arcsec$/pixel, this shift should be detectable for many foreground stars.}

\begin{figure}[ht]

\begin{center}

\includegraphics[width=0.45\textwidth]{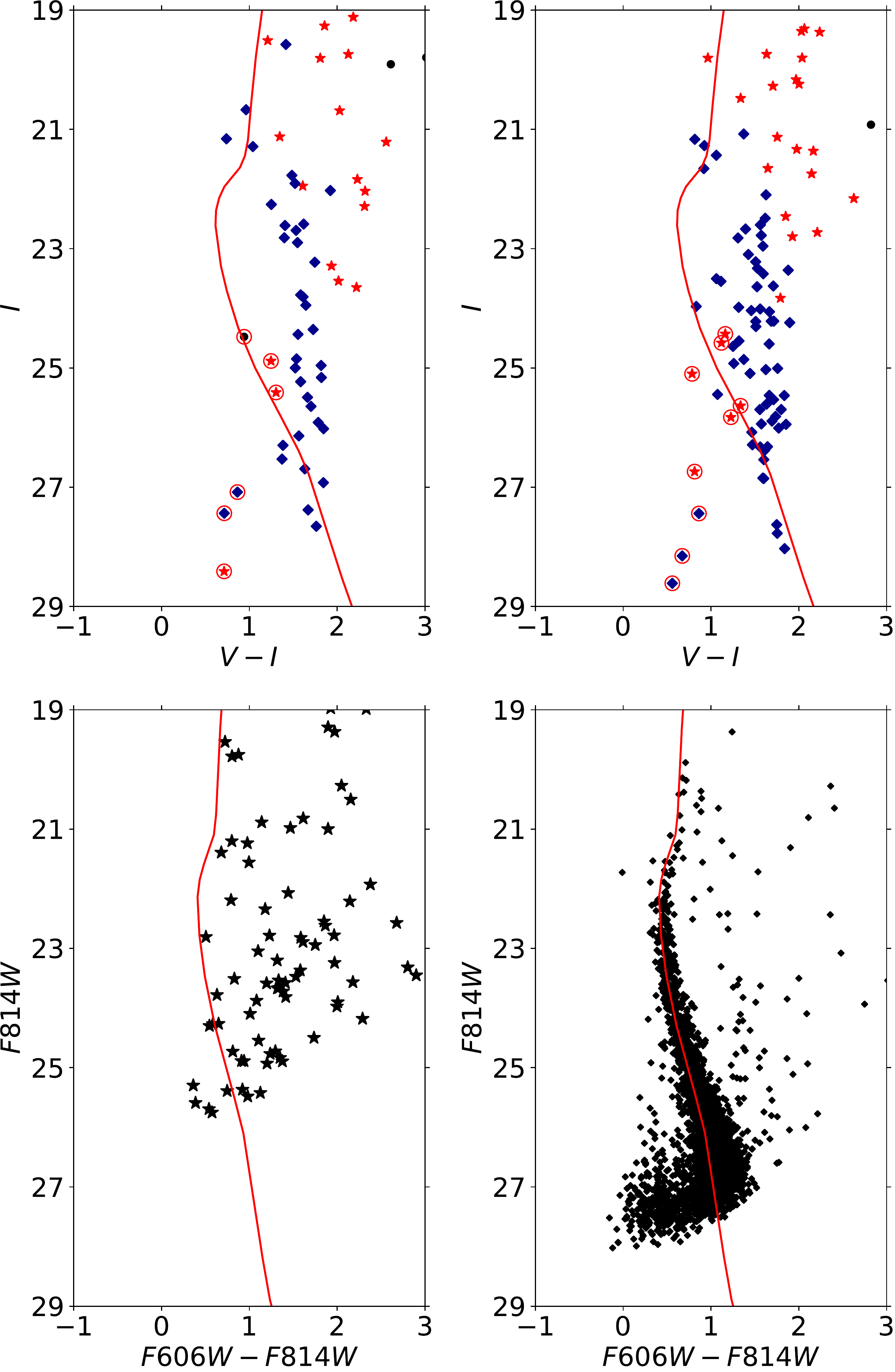}

\caption{{\it Upper panels:\/} The {predicted distributions of Milky Way stars in color-magnitude space from the Besan{\c c}on model, for lines-of-sight corresponding to Off Field 2 (left) and \boo\ (right).   The output from the Besan{\c c}on model is plotted in Johnsons-Cousins V and I.  Halo stars are plotted as dark blue diamonds, thick-disk stars as red stars, and thin-disk stars as black circles. White dwarfs are indicated by a red circle around the symbol.}
    {\it Lower panels:\/} {Color-magnitude diagrams from the DAOPHOT photometry for Off Field 2 (left) and \boo\ (right).}
{     The red curve in all panels is a 13~Gyr isochrone of metallicity equal to the peak of the observed metallicity distribution of \boo,  ${\rm [M/H]} = -2.2$, appropriately adjusted for dust in that line-of-sight and an assumed distance of  62~kpc.}}

\label{fig:contcomp}

\end{center}

\end{figure}

The {distribution of stellar sources in the off-field is shown in the lower left panel of Figure~\ref{fig:contcomp}, with the predictions from the Besan{\c c}on model in the upper left panel. The lower latitude of this field is reflected in the higher number of stars predicted, but again most sources lie far from the locus of the main population of \boo. It should be noted that while the off-field data do not reach as faint as the \boo\ field data, {we use the off-field to understand the contribution of stars that are not members of \boo. The majority of stellar contaminants are brighter than $\sim 25.5$ (as seen in Figure \ref{fig:contcomp}), and the cloud of blue sources around 27th magnitude are likely faint background galaxies and a few Galactic white dwarfs, discussed in Section \ref{sec:galaxies}. The relative importance of stellar contaminants is higher above a magnitude of $\sim 25.5$, motivating our use of a shallower off-field.}}

\subsubsection{Faint Galaxy Contamination}\label{sec:galaxies}

{The faint blue sources (colors $\simlt 0.5$, magnitudes $\simgt 26$) in the color-magnitude diagrams are likely to be predominantly high-redshift, marginally resolved galaxies, plus a few white dwarfs in  the Milky Way (they are too bright to be white dwarfs in \boo).  The analysis by \cite{bedin}  supports the idea that there is a population of blue galaxies  that is not distinguishable from stars in ACS/WFC images, using standard techniques. Those authors  re-analyzed the data from the Hubble Ultra Deep Field \citep{hudf}, applied their stellar photometric pipeline and cleaning and found  that faint blue  galaxies  made it through the processing and occupied a location on the color-magnitude diagram that is overlapping, but not identical to, the region that we see populated (see their Figure~9). {Though} we adopted a different photometric pipeline and cleaning process than did \cite{bedin}, it is {likely} that the faint, blue sources of our analysis  are also faint galaxies. {This idea is further supported by inspecting the \textit{sharp} distribution of the faint blue detections, which one would expect to peak at positive values for unresolved galaxies (a wider PSF than a true stellar object). Indeed, in both F606W and F814W, the \textit{sharp} distribution is peaked at positive values, but still small enough to have passed through the cleaning process ($ \simlt 0.2 $, versus $\sim$ 0 for the remaining sources in our catalog). Happily, the projected location of faint blue galaxies in color-magnitude space is bluer than low-mass stars of the metallicity and distance of \boo, as evidenced by the isochrones in Figure \ref{fig:isochronecmd} and by the discussion in \cite{bedin}. This separation in color-magnitude space leads us to conclude that these faint blue galaxies do not severely contaminate our sample. The bulk of these sources were given low weight (see Figure~\ref{fig:weightedcmd}) and so would be removed by our membership scheme.}

\section{Discussion and Conclusions} \label{sec:discussion}

The core goal of this work was to produce photometry for stars {in \boo, reaching as faint as possible down the main sequence, with $0.3 \msun$ being the target limiting mass. As discussed above, in Section~\ref{sec:artstar}, the ACS/WFC DAOPHOT photometry has a $50\%$ completeness limit of 27.4 Vega magnitude in F814W, and 28.2 Vega magnitude in F606W. These correspond to absolute magnitudes of +8.41 in F814W and +9.20 in F606W (using the distance modulus given in Table~\ref{tab:boo-param} and dust extinction from Section~\ref{sec:dust}).}

{The \cite{dartmouth} {models} we used in Section \ref{sec:CMD} translate the F814W limit to  $0.29 \cal M_\odot$ for $[M/H] = -1.7$ and  $0.26 \cal M_\odot$ for $[M/H] = -2.5$.}{\citet{cassisi} calculated models for very low mass, low metallicity stars that we can use to obtain an independent estimate of the limiting mass.  Their lowest metallicity  models had $Z = 0.0002$ and an age of 10~Gyr (see their Table~1). This age is younger than that usually derived for \boo, but there should be negligible effect on the predicted luminosities of the low-mass stars of interest. {These models provide a low-mass limit of between  $0.3 \cal M_\odot$ and $0.25 \cal M_\odot$ for our F814W $50\%$ completeness limit.}   We therefore conclude that we did, indeed, produce photometry that reaches stars of $0.3 \cal M_\odot$ or lower, achieving our goal.

This  low-mass limit of $\simlt 0.3 \msun$ allows us to enter the regime discussed by \cite{elbadry}, where potential deviations from the Milky Way IMF can be discerned. The second paper in this series, {Filion et al.~(2021a.), uses the results of this paper} in a Bayesian analysis (allowing unresolved binaries to be incorporated) of the low-mass end of the stellar IMF of \boo. The high precision UVIS photometry will be employed in paper III {Filion et al.~(2021b.),} alongside the ACS/WFC photometry, to constrain the star-formation history of \boo. {These future studies will be aided by  proper motions derived from our observations and those of GO-12549, which will be used to identify non-member stars.}

\section{Acknowledgements} \label{sec:acknowledgements}
{CF and RFGW are grateful for support} through the generosity of Eric and Wendy Schmidt, by recommendation of the Schmidt Futures program. RFGW thanks her sister, Katherine Barber, for her support  of this research. {We thank Aaron Dotter for his help with the Dartmouth stellar evolution models.} This research is based on observations made with the NASA/ESA Hubble Space Telescope obtained from the Space Telescope Science Institute, which is operated by the Association of Universities for Research in Astronomy, Inc., under NASA contract NAS 5–26555. These observations are associated with programs GO-15317, GO-12549, GO-13393. Support for this analysis of data from program 15317 was provided by NASA through a grant from the Space Telescope Science Institute, which is operated by the Association of Universities for Research in Astronomy, Inc., under NASA contract NAS 5–26555. This work has made use of the SVO Filter Profile Service \url{http://svo2.cab.inta-csic.es/theory/fps/} supported from the Spanish MINECO through grant AYA2017-84089. 
This research has also made use of the VizieR catalogue access tool, CDS, Strasbourg, France (DOI : 10.26093/cds/vizier). The original description of the VizieR service was published in \cite{vizier}. Additionally, this work made use of the SciServer science platform ({\tt www.sciserver.org}). SciServer is a collaborative research environment for large-scale data-driven science. It is being developed at, and administered by, the Institute for Data Intensive Engineering and Science at Johns Hopkins University. SciServer is funded by the National Science Foundation through the Data Infrastructure Building Blocks program and others, as well as by the Alfred P. Sloan Foundation and the Gordon and Betty Moore Foundation.

\facility{HST (ACS, WFC3)}

\software{astroML \citep{astroML},\ 
  Astropy (\citealt{astropy-13},\  \citealt{astropy-18}),\ 
  DrizzlePac \citep{drizzle},\  
  IRAF \citep{iraf1,iraf2},\ 
  Matplotlib \citep{hunter},\  
  pandas \citep{pandas},\ 
  TOPCAT \citep{taylor}}





\end{document}